\documentclass[a4paper,fleqn]{cas-sc}

\usepackage[utf8]{inputenc}
\usepackage{subfig}
\usepackage{floatpag}
\usepackage{float}
\usepackage{graphicx}
\graphicspath{ {./images/} }
\usepackage{caption}
\usepackage{subcaption}
\usepackage[sort,numbers]{natbib}
\usepackage{hyperref}
\usepackage[capitalise]{cleveref}
\usepackage{mathtools}
\usepackage[export]{adjustbox}
\usepackage[section]{placeins}
\def\tsc#1{\csdef{#1}{\textsc{\lowercase{#1}}\xspace}}
\tsc{WGM}
\tsc{QE}

\usepackage{float}
\begin{document}
\let\WriteBookmarks\relax
\def\floatpagepagefraction{1}
\def\textpagefraction{.001}
\newcommand{\overbar}[1]{\mkern 1.5mu\overline{\mkern-1.5mu#1\mkern-1.5mu}\mkern 1.5mu}
\shorttitle{Shock capturing meshless method}    

\shortauthors{Satyaprasad et al.}

\title [mode = title]{A meshless geometric conservation weighted least square method for solving the shallow water equations}  



%

\author[1]{D. Satyaprasad}[type=author]





\credit{Conceptualization and implementation of the numerical method and computation, Writing manuscript}

\affiliation[1]{organization={Department of Mathematics},
            addressline={IIT Madras}, 
            city={Chennai},
            citysep={}, 
            postcode={600036}, 
            state={Tamilnadu},
            country={India}}

\author[2]{Soumendra Nath Kuiry}[type=author]




\credit{Discussion on the numerical results, Revising the manuscript, Supervision}
\affiliation[2]{organization={Department of Civil Engineering},
            addressline={IIT Madras}, 
            city={Chennai},
          citysep={}, 
            postcode={600036}, 
            state={Tamilnadu},
            country={India}}
\cormark[2]
\cortext[2]{Corresponding author.
}
\ead{snkuiry@civil.iitm.ac.in}


\author[1]{S. Sundar}[type=author]
\credit{Discussion on the numerical results, Supervision}





\begin{abstract}
The shallow water equations are numerically solved to simulate free surface flows. The convective flux terms in the shallow water equations need to be discretized using a Riemann solver to capture shocks and discontinuity for certain flow situations such as hydraulic jump, dam-break wave propagation or bore wave propagation, levee-breaching flows, etc. The approximate Riemann solver can capture shocks and is popular for studying open-channel flow dynamics with traditional mesh-based numerical methods. Though meshless methods can work on highly irregular geometry without involving the complex mesh generation procedure, the shock-capturing capability has not been implemented, especially for solving open-channel flows. Therefore, we have proposed a numerical method, namely, a shock-capturing meshless geometric conservation weighted least square (GC-WLS) method for solving the shallow water equations. The HLL (Harten-Lax-Van Leer) Riemann solver is implemented within the framework of the proposed meshless method. The spatial derivatives in the shallow water equations and the reconstruction of conservative variables for high-order accuracy are computed using the GC-WLS method. The proposed meshless method is tested for various numerically challenging open-channel flow problems, including analytical, laboratory experiments, and a large-scale physical model study on dam-break event.
\end{abstract}


\begin{highlights}
\item A shock-capturing meshless method is presented for solving the 2D shallow water equations on a highly variable topography.
\item The meshless coefficients are determined by applying the geometric conservation law and first-order consistency through the Lagrange multiplier method, ensuring discrete local conservation within the approach.
\item The HLL Riemann solver is used in the framework of the meshless method to capture shocks and flow discontinuities.

\end{highlights}


\maketitle

\section{Introduction}
In computational fluid dynamics (CFD), various flow phenomena like flood wave propagation, dam-break scenarios, hydraulic jumps, and more complex flows can be effectively described by the two-dimensional (2D) shallow water equations. These equations are utilized to simulate 2D flows featuring discontinuities and shocks numerically. Numerous numerical techniques have been developed over time to solve these equations, including the finite difference method (FDM) \cite{fennema1990explicit,molls1995depth}, finite element method (FEM) \cite{FEM2008}, and finite volume method (FVM) \cite{Yoon2004,Kuiry2008231,EYMARD2000713}, among others. However, whether structured or unstructured, conventional mesh-based methods face limitations in handling intricate geometries and generating corresponding flow solvers, requiring substantial effort to represent minute topographical features in the computational mesh. To address these challenges, a class of methods referred to as meshless, gridless, meshfree, or particle methods has gained popularity in recent years for simulating, especially compressible Euler equations. These methods offer inherent flexibility in dealing with complex geometries. Notable meshless methods include Smoothed Particle Hydrodynamics (SPH) \cite{Monaghan_SPH}, Finite Pointset Method (FPM) \cite{Kuhnert_FPM_5,Kuhnert_FPM_6}, Element-free Galerkin (EFG) \cite{BELYTSCHKO_EFG} method, Generalized Finite Difference Method (GFDM) \cite{GFDMSWE}, and discrete mixed subdomain least squares method \cite{Fazli_Malidareh}, to name a few. However, meshless techniques are relatively novel in free surface flow simulations, particularly in capturing shocks and discontinuities by solving the shallow water equations, which remain under-explored in literature. 

Among meshfree methods, SPH is the earliest developed and commonly used meshfree method. Originally created for astrophysical fluid dynamics \cite{Monaghan_SPH}, the SPH was later adapted by Monaghan \cite{Monaghan} for hydrodynamic modeling, particularly for solving the Navier-Stokes equations for incompressible free-surface flows. A significant challenge in the SPH lies in incorporating boundary conditions essential for solving real-world flow dynamics needed by the shallow water equations. Subsequently, the EFG method, introduced by Belytschko et al. \cite{BELYTSCHKO_EFG}., emerged as an extension of the Diffuse Element (DE) method proposed by Nayroles et al.  \cite{Nayroles1992}. In the EFG method, the generalized Moving Least Square (MLS) interpolation is employed to define the local approximation of the spatial derivatives. Another notable meshfree approach is the FPM, developed by Tiwari and Kuhnert \cite{Kuhnert_FPM_5,Kuhnert_FPM_6}, which finds extensive applications in various fluid dynamics problems \cite{Kuhnert_FPM_3,Kuhnert_FPM_4,Kuhnert_FPM_1,Kuhnert_FPM_2,Kuhnert_FPM_7,Kuhnert_FPM_8,Kuhnert_FPM_9}. The FPM utilizes a Weighted Least Square (WLS) method to approximate the spatial derivatives across the computational domain. It serves as a Lagrangian model particularly suited for fluid problems involving rapidly changing flow domains over time \cite{Kuhnert_FPM_5, Kuhnert_FPM_1}, offering simplicity in implementation and setting up necessary boundary conditions \cite{Kuhnert_FPM_5,Kuhnert_FPM_6}.

However, contrary to their names, most mesh-free methods do not entirely eliminate the need for meshes. Although these methods do not rely on a predefined mesh for domain discretization, many ofthem still incorporate a readily generable mesh. In such cases, the mesh quality does not significantly impact the solution, ensuring that the method's effectiveness is not overly contingent on the mesh quality \cite{Liu_meshfreemethods}. Nonetheless, several noteworthy meshless solvers for shallow water problems have emerged over time, including the GFDM by Li and Fan \cite{GFDMSWE}, the FPM by Buachart et al. \cite{SWEFPM}, and the Radial Basis Function (RBF) method by Chaabelasri \cite{RBFDamBreak}. However, applying GFDM to shallow water equations often leads to oscillations near discontinuities, particularly evident in one-dimensional (1D) dam-break scenarios, as highlighted by Li and Fan \cite{GFDMSWE}. Similarly, Buachart et al.'s \cite{SWEFPM} FPM encounters challenges when dealing with intricate source terms in shallow water equations. On the other hand, the RBF-based meshless approach \cite{RBFDamBreak} for solving shallow water equations incorporates artificial viscosity to capture shocks and has been evaluated only against regularly distributed points. Furthermore, these meshless methods for shallow water equations have not undergone thorough verification across analytical, experimental, and real-life flow scenarios to the extent seen in the traditional mesh-based finite volume methods (FVMs). Nevertheless, the scientific community doubts these meshless methods primarily due to their lack of formal conservation property. Based on our understanding, the existing mesh-free schemes do not preserve conservation at the discrete level because of their local nature, except in very limited situations, like those involving uniform point distributions that yield trivial meshes. On the other hand, the FVMs exhibit the characteristic of local discrete conservation of fluxes between adjacent control volumes, facilitated by their predefined geometry and tessellation of the control volumes. The FVM discretization for shallow water equations often employs the Riemann solvers to precisely capture shocks and flow discontinuities, eliminating the need for artificial viscosity. Thus, integrating Riemann solvers into the meshless methods can augment their ability to capture crucial physical aspects of shallow water flow dynamics, including discontinuities, shocks, sudden variations in bed topography, and flow variables like depth and velocity. A comprehensive review of existing literature by the authors reveals a gap in research concerning the integration of local discrete conservation and the Riemann solvers in meshless methods, particularly in simulating discontinuous flows and flows over highly variable terrain. Therefore, adopting a conservative numerical method and the HLL (Harten-Lax-Van Leer) Riemann solver proposed by Harten et al. \cite{HLL} with a two-wave model resolving three constant states can enhance the effectiveness of meshless methods in addressing practical shallow water flow problems. Satyaprasad et al. \cite{D_Satyaprasad} developed a meshless model for solving the 1D Saint-Venant equations. The study reported that the method is highly accurate and does not diffuse at the wavefront. However, a 1D meshless model cannot solve most practical open-channel flow problems, so a 2D model is necessary.

In this paper, we have developed a meshless geometric conservation technique to numerically solve the 2D shallow water equations aimed at simulating flows in non-prismatic open channels characterized by sudden changes in topography. In this approach, we have integrated the HLL approximate Riemann solver into the meshless framework to handle shocks and discontinuities effectively. The geometric conservation least square (GC-LSM)\cite{GCWLS} is adopted to approximate the spatial derivatives. The proposed numerical scheme is well-balanced due to the adoption of the shallow water equations in the form given by Ying and Wang \cite{Ying_Wang_2009}, which automatically incorporates the surface gradient method of Zhou et al.\cite{ZHOU20011}. A collection of irregularly scattered points describes the physical domain. A local cluster of nearby points, termed satellites, is identified for each of these points. Local approximations of the spatial derivatives are estimated using the Taylor-series based WLS method, constrained by the geometric conservation law with first-order consistency. In our implementation, we adopt a Gaussian weight function, as outlined in the FPM method by Tiwari and Kuhnert \cite{Kuhnert_FPM_6}. We call this geometric conservation weighted least square method (GC-WLS). The meshless coefficients in GC-WLS define the approximation of the spatial derivatives. The convective fluxes are determined by employing the HLL Riemann solver to capture discontinuous flow patterns. Consequently, the HLL Riemann solver uses meshless coefficients to approximate numerical fluxes. The source term's water surface gradient is treated by averaging the water surface levels at the midpoints of the central and satellite points of the local cluster, utilizing the piece-wise linearly reconstructed variables and meshless coefficients. Similarly, driving forces and fluxes are computed using the reconstructed variables, ensuring consistency and well-balance in the solution scheme. By employing the HLL Riemann solver and maintaining consistent discretization, our proposed meshless method is adept at handling discontinuous and shock flows over varying bed topography. The performance of the proposed meshfree method is verified by solving various analytical test cases, including the 1D dam-break flow on a wet and dry horizontal bed, and moving shorelines in a 2D frictional parabolic bowl. The proposed meshless method is then validated by simulating laboratory experiments on dam break wave propagation over a triangular hump and a real-life Malpasset dam-break event. The results show that the proposed shock-capturing meshless method is highly accurate and does not diffuse at the wavefront. The meshless method, incorporating the HLL approximate Riemann solver, prevents the occurrence of oscillations at the wavefront and effectively manages intricate source terms. This sets the proposed shock-capturing meshless method apart from the established meshless methods, especially GFDM and FPM (Buachart et al., 2013; Li and Fan, 2017). Additionally, the proposed method eliminates the need for artificial viscosity to mitigate numerical oscillations that were noticeable in prior implementations (e.g., Chaabelasri, 2018).

\section{Governing equations}
The 2D shallow water equations are used as the governing equations for the proposed model. The equations can be derived by integrating the Navier-Stokes equations over the flow depth and can be written in a conservative form as suggested by Ying and Wang \cite{Ying_Wang_2009} below.

\begin{equation} \label{goveqn}
\frac{\partial \boldsymbol{U}}{\partial t}+\frac{\partial \boldsymbol{G}}{\partial x}+\frac{\partial \boldsymbol{H}}{\partial y}=\boldsymbol{S}
\end{equation}
where $\mathbf{U}$ represents the vector of conservative variables, $\mathbf{G}$ and $\mathbf{H}$ define flux vectors in the $x$ and $y$ directions, and $\mathbf{S}$ is the source term vector defined as
\begin{equation*}
\mathbf{U}=\left[\begin{array}{l}
h \\
hu \\
hv
\end{array}\right], \quad \mathbf{G}=\left[\begin{array}{l}
hu \\
hu^2\\
huv
\end{array}\right], \quad \mathbf{H}=\left[\begin{array}{l}
hv \\
huv\\
hv^2
\end{array}\right], 
\end{equation*}

\begin{equation} \label{sourceTerm}
\mathbf{S}=\mathbf{S}_b + \mathbf{S}_f=\begin{bmatrix}
0 \\
-gh\displaystyle\frac{\strut\partial Z}{\strut\partial x}\\
-gh\displaystyle\frac{\strut\partial Z}{\strut\partial y}
\end{bmatrix} + \begin{bmatrix}
0 \\
-C_f u \sqrt{u^2+v^2}\\
-C_f v \sqrt{u^2+v^2}\\
\end{bmatrix}
\end{equation}

where $h$ is water depth, and $u$ and $v$ are the depth-averaged velocity components in the $x$ and $y$ directions, respectively. $g$ is the gravitational acceleration with a value $9.81$ $m/s^2$, $Z$ is the water surface level which equals to $h + z_b$, herein $z_b$ is the bed elevation from a datum, and $C_f$ is the bed roughness coefficient controlled by Manning's roughness coefficient $n$, represented as $C_f = gn^2/h^{1/3}$. 


\section{Numerical method}
The meshless numerical methods represent a category of computational approaches to solve partial differential equations (PDEs) without needing a predefined mesh or grid structure. These techniques offer distinct advantages, especially in scenarios featuring intricate geometries, dynamic boundaries, or substantial grid deformations, where creating a structured or unstructured mesh is either impractical or computationally intensive. In the case of hydraulics, erosion or deposition by sediment transport can demand an adaptive mesh due to large deformation in the bed topography. In such a case, a fixed mesh is largely deformed, and an adaptive mesh can be computationally intensive due to the dynamic nature of the process. However, the meshless methods can be applied to moving or fixed geometries without paying much attention to mesh quality. Because of the inherent characteristics of the fixed bed geometry present in open channel flows (i.e., without considering bed deformation due to sediment transport), we have opted to use the Eulerian framework for solving the 2D shallow water equations (\cref{goveqn}). In the method suggested herein, the points initially distributed are not required to be maintained throughout the simulation.

This study presents a meshless numerical technique utilizing the WLS method for solving the 2D shallow water equations (\cref{goveqn}). The proposed method delineates the physical domain by a collection of irregularly distributed finite points, each possessing fluid attributes like depth, velocity, momentum, etc. We identify a nearby cluster of neighbouring points for every considered point, called satellites. Local approximations of the spatial derivatives are estimated from the Taylor series using the WLS method constrained with geometric conservation law and first-order consistency. The approximation of these spatial derivatives is described by meshless coefficients. The convective fluxes in equation (\cref{goveqn}) are computed utilizing the HLL Riemann solver that captures flow discontinuity. Including source terms is crucial for addressing real-world open-channel flow dynamics. To prevent numerical discrepancies between bed slope and convective flux terms, we adhere to the SGM approach \cite{ZHOU20011}. The gradients in source terms are also managed using meshless coefficients to ensure equilibrium between convective fluxes and source terms.

\subsection{WLS Method}
In various real-world scenarios, selecting mesh-based numerical schemes significantly impacts solution accuracy. Nonetheless, their precision might diminish with inadequately distributed grid points. The approach outlined here doesn't rely on regular grids to estimate spatial derivatives of a function. Instead, it utilizes a cloud of grid points, offering a meshless alternative. In this section, we introduce the fundamental theory of the WLS method for approximating spatial derivatives of a function.\\

Let $\Omega \subset \mathbb{R}^2$ be a domain containing $N$ number of scattered points and $f(x,y)$ be a scalar function and $f(x_i, y_i)$ be function values at $(x_i, y_i) \in \Omega $ for $i=1,2, \cdots N.$ These points are the numerical grid points, which are not necessarily being connected. For every point, $(x_i, y_i)$, consider the neighbouring cloud of points $C_i$, in which the point $(x_i, y_i)$ is the centre, and the other points are called the satellites. Let $M_i$ be the total number of satellites in the cloud $C_i$.\\

\begin{figure}[h]
     \centering
         \includegraphics[scale=0.9]{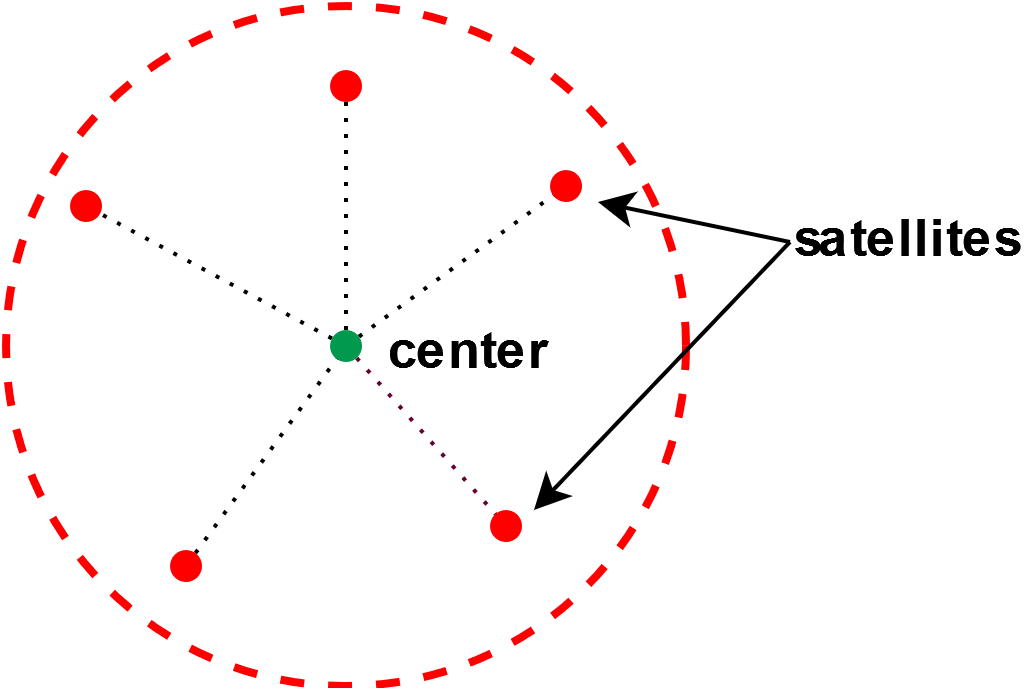}
        \caption{A typical meshless cloud of points}
        \label{fig:cloudOfPoints}
\end{figure}

The coordinate difference between the satellite $(x_j, y_j)$ and the centre $(x_i, y_i)$ can be expressed as 
\begin{equation}
    h_{ij} = x_j - x_i, \mbox{~~~} k_{ij} = y_j - y_i
\end{equation}
Let the vector $\vec{r}_i^j = (h_{ij}, k_{ij})$ starts from $(x_i, y_i)$ to $(x_j, y_j)$, its length is 
\begin{equation}
    r_{i}^{j}=\sqrt{\displaystyle h_{ij}^2 + k_{ij}^2}
\end{equation}
 and the reference radius of the cloud is defined as 
\begin{equation}
    R_{i}=\max\limits_{(x_j, y_j) \in C_i} r_{i}^{j}
\end{equation}

To approximate the spatial derivatives, the WLS method \cite{Kuhnert_FPM_6} is applied in the present work like the earlier studies of meshfree methods for solving flow problems \cite{Kuhnert_FPM_5, Batina, wlsmethod_zhihua}. We consider a Gaussian weight function of the form 
\begin{equation}
    w\left(\boldsymbol{x}_{i}-\boldsymbol{x} ; R_{i}\right)=\left\{\begin{array}{cc}
\exp \left(-\gamma \frac{\left\|\boldsymbol{x}_{i}-\boldsymbol{x}\right\|^{2}}{R_{i}^{2}}\right), & \quad \text { if } \frac{\left\|\boldsymbol{x}_{i}-\boldsymbol{x}\right\|}{R_{i}} \leq 1 \\
0, & \text { otherwise }
\end{array}\right.
\end{equation}
where $\gamma$ is a constant chosen in the range between 2 to 7.\\
Let $\boldsymbol{x} = (x,y)$ and consider any differentiable function $f(\boldsymbol{x})$ in a given small domain $\Omega_i$, the Taylor series about a point $\boldsymbol{x}_i$ can be expressed in the following form

\begin{equation} \label{eq:1}
f(\boldsymbol{x}) = f(\boldsymbol{x}_i) + \nabla f_i \cdot (\boldsymbol{x} - \boldsymbol{x}_i) + \frac{1}{2} (\boldsymbol{x} - \boldsymbol{x}_i)^T D_i (\boldsymbol{x} - \boldsymbol{x}_i) + O(\boldsymbol{l}^3)
\end{equation}
where $\boldsymbol{l} = \boldsymbol{x} - \boldsymbol{x}_i$ is the error in the Taylor's expansion. The coefficients $a = [\nabla f_i, D_i]^T$ represent the partial derivatives of the function $f$ at $\boldsymbol{x}_i$. By keeping the terms up to  second-order, the approximate function value at point $\boldsymbol{x}_j$ is obtained as

\begin{equation} \label{eq:3}
    \tilde{f}(\boldsymbol{x_j}) = f(\boldsymbol{x}_i) + \nabla f_i \cdot (\boldsymbol{x_j} - \boldsymbol{x}_i) + \frac{1}{2} (\boldsymbol{x_j} - \boldsymbol{x}_i)^T D_i (\boldsymbol{x_j} - \boldsymbol{x}_i)
\end{equation} 
As shallow water equations (\ref{goveqn}) have first-order partial derivatives, it is reasonably enough to maintain up to first-order terms in the equation (\ref{eq:3}). Hence, the approximate value is
\begin{equation}
    \tilde{f}(\boldsymbol{x_j}) = f(\boldsymbol{x}_i) + \nabla f_i \cdot (\boldsymbol{x_j} - \boldsymbol{x}_i) 
\end{equation} 
and the error between the exact and approximate value is 
\begin{equation} \label{eq:2}
    \boldsymbol{e}_j = f(\boldsymbol{x}_j) - \tilde{f}(\boldsymbol{x_j}) = f(\boldsymbol{x}_j) - (f(\boldsymbol{x}_i) + \nabla f_i \cdot (\boldsymbol{x_j} - \boldsymbol{x}_i))
\end{equation}
The coefficients $\boldsymbol{a} = \nabla f_i $ in Eq. (\ref{eq:2}) are computed by minimizing the following norm over the neighbouring cloud $C_i$

\begin{equation}\label{eq:minima}
    \Phi = \sum_{j = 1}^{M_i} w_j \boldsymbol{e}_j^2
\end{equation}
and minimization of $\Phi$ is given by $\displaystyle \frac{\partial \Phi}{\partial \boldsymbol{a}} = 0$. This yields
\begin{equation}
    A\boldsymbol{a} = \boldsymbol{b}
\end{equation}
where 
\begin{equation}
    A = \left[ \begin{array}{ll}
    \displaystyle \sum_{j=1}^{M_i} w_j h_{ij}^2 & \displaystyle \sum_{j=1}^{M_i} w_j h_{ij}k_{ij} \\
    &\\
    \displaystyle \sum_{j=1}^{M_i} w_j h_{ij}k_{ij} & \displaystyle \sum_{j=1}^{M_i} w_j k_{ij}^2 \\
    \end{array}\right]
\end{equation}

\begin{equation}
    \boldsymbol{a} = \left[\begin{array}{c}
         a_1  \\
         \\
         a_2  \\
    \end{array}\right] \hspace{0.5cm}
    \boldsymbol{b} = \left[\begin{array}{c}
        \displaystyle \sum_{j=1}^{M_i} w_j(f_j - f_i)h_{ij}\\
        \\
        \displaystyle \sum_{j=1}^{M_i} w_j(f_j - f_i)k_{ij}\\
    \end{array}\right]
\end{equation}
The above system of equations can be solved as
\begin{equation} \label{Eq-solution}
    \boldsymbol{a} = A^{-1}\boldsymbol{b}
\end{equation}
The solution can be written into a linear combination of the function values at different points
\begin{equation}
    a_1 = \left.\frac{\partial f}{\partial x}\right|_i = \sum_{j=1}^{M_i}\alpha_j(f_j - f_i) \mbox{ ~~ },
    a_2 = \left.\frac{\partial f}{\partial y}\right|_i = \sum_{j=1}^{M_i}\beta_j(f_j - f_i)
\end{equation}
where $\alpha_j$ and $\beta_j$ are computed from Eq. (\ref{Eq-solution}). The spatial derivatives can also be estimated using the midpoint $ij$ between $i$ and $j$ as
\begin{equation} \label{spatialderivative}
    a_1 = \left.\frac{\partial f}{\partial x}\right|_i = \sum_{j=1}^{M_i} \alpha_{ij}(f_{ij} - f_i) \mbox{ ~~ }
    a_2 = \left.\frac{\partial f}{\partial y}\right|_i = \sum_{j=1}^{M_i} \beta_{ij}(f_{ij} - f_i)
\end{equation}
where $\alpha_{ij}, \beta_{ij}$ are meshless coefficients, $f_{ij}$ is estimated at the midpoint between $i$ and $j$ 

\subsection{Approximation of Derivatives with Geometric Conservation Weighted Least Square method (GC-WLS)}
\begin{figure}[hbt!]
     \centering
     \subfloat[]{
     \begin{minipage}[b]{0.5\textwidth}
         \centering
         \includegraphics[width=0.8\textwidth]{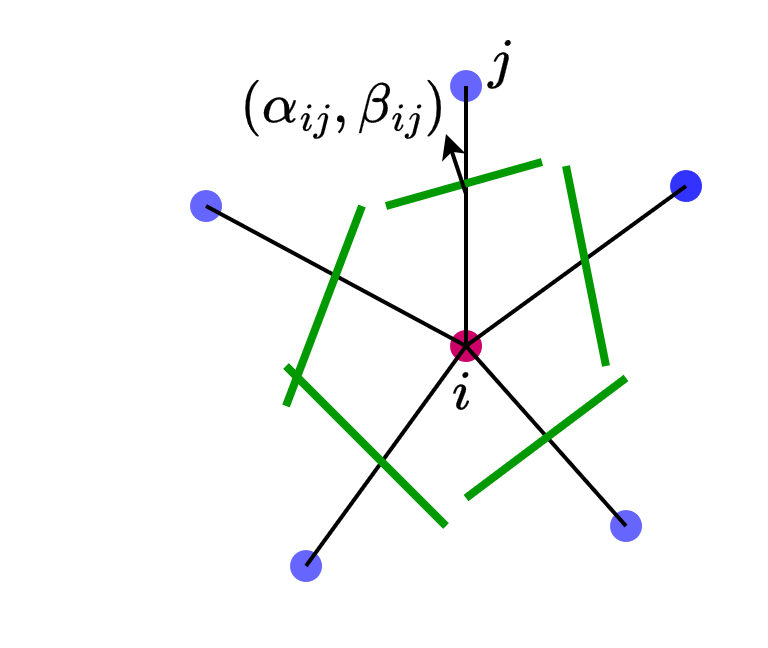}
     \end{minipage}}
     \subfloat[]{
     \begin{minipage}[b]{0.5\textwidth}
         \centering
         \includegraphics[width=0.7\textwidth]{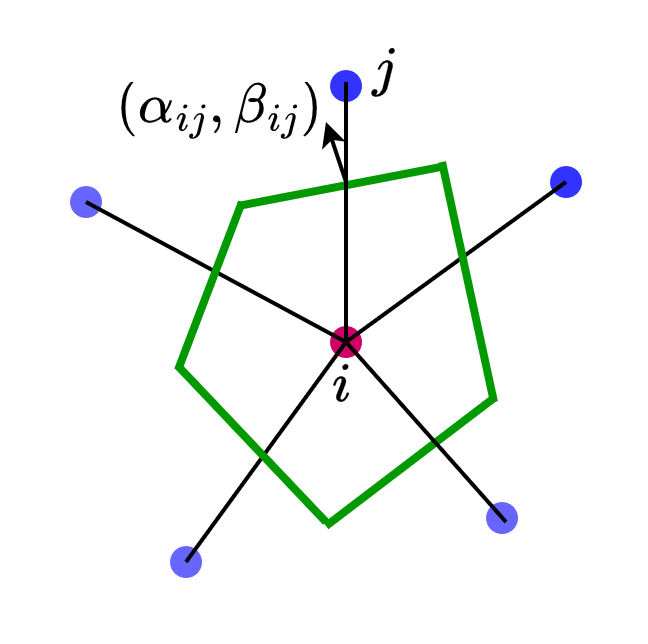}
     \end{minipage}}
        \caption{Geometric representation of virtual cell: (a) WLS (b) GC-WLS}
        \label{fig:GC}
\end{figure}

In traditional mesh-based methods, discrete local conservation is achievable using a conservative numerical scheme. This is ensured by the stipulation that mandates each cell to possess a specific volume, fully enclosed by its faces without any overlapping \cite{conservation_meshfree}. In order to achieve discrete local conservation in meshless methods, the geometry of the virtual cell between $i$ and $j$, defined by meshless coefficients, should be fully enclosed by the virtual faces between the points $i$ and $j$. Figure \ref{fig:GC} shows the typical geometric shapes of virtual cells for WLS and GC-WLS. To achieve such meshless coefficients, we multiply the first-order terms of \cref{eq:1} by $\alpha_{j}$, $\beta_{j}$ and then sum over 1 to $M_i$ as written below.

\begin{equation}\label{eq:taylor_1}
    \displaystyle \sum_{j=1}^{M_i} \alpha_{j}{f}(\boldsymbol{x_j}) \approx \left(\sum_{j=1}^{M_i} \alpha_{j}\right)f(\boldsymbol{x_i}) + \left(\sum_{j=1}^{M_i} \alpha_{j}h_{ij}\right) \left.\frac{\partial f}{\partial x}\right|_i + \left(\sum_{j=1}^{M_i} \alpha_{j}k_{ij}\right) \left.\frac{\partial f}{\partial y}\right|_i 
\end{equation}

\begin{equation}\label{eq:taylor_2}
    \displaystyle \sum_{j=1}^{M_i} \beta_{j}{f}(\boldsymbol{x_j}) \approx \left(\sum_{j=1}^{M_i} \beta_{j}\right)f(\boldsymbol{x_i}) + \left(\sum_{j=1}^{M_i} \beta_{j}h_{ij}\right) \left.\frac{\partial f}{\partial x}\right|_i + \left(\sum_{j=1}^{M_i} \beta_{j}k_{ij}\right) \left.\frac{\partial f}{\partial y}\right|_i 
\end{equation}
The first-order spatial derivatives of the function $f$ can be computed by considering the following linear combination formats
\begin{equation}\label{eq:partial_der}
    \left.\frac{\partial f}{\partial x}\right|_i = \sum_{j=1}^{M_i}\alpha_{j} (f_j -f_i) \mbox{ ~~ }, and
    \left.\frac{\partial f}{\partial y}\right|_i = \sum_{j=1}^{M_i}\beta_{j} (f_j -f_i) 
\end{equation}
The expressions in \cref{eq:partial_der} can be approximated from \cref{eq:taylor_1} and \cref{eq:taylor_2} if the following conditions are satisfied.

Geometric conservation law:
\begin{equation}\label{eq:geometric}
\begin{array}{ll}
    \displaystyle \sum_{j=1}^{M_i} \alpha_{j} = 0, &\displaystyle \sum_{j=1}^{M_i} \beta_{j} = 0 \\
  \end{array}  
\end{equation}

First-order consistency:
\begin{equation}\label{eq:firstorder}
\begin{array}{ll}
    \displaystyle \sum_{j=1}^{M_i} \alpha_{j}h_{ij} = 1, & \displaystyle\sum_{j=1}^{M_i} \beta_{j}h_{ij} = 0 \\
    &\\
    \displaystyle \sum_{j=1}^{M_i} \alpha_{j}k_{ij} = 0, & \displaystyle\sum_{j=1}^{M_i} \beta_{j}k_{ij} = 1 \\
  \end{array}    
\end{equation}

The meshless method satisfies the conservation law if the meshless coefficients $\alpha_{j}$ and $\beta_{j}$ satisfy the geometric conservation law and the first-order consistency \cite{GCWLS}. For such meshless coefficients, the Lagrange multiplier method is used to find the minima of \cref{eq:minima} with the constrained equations \cref{eq:geometric} and \cref{eq:firstorder}. The Lagrange function can then be defined as
\begin{equation}
    L = \Phi_i + \mu_1\sum_{j=1}^{M_i} \alpha_{j} + \mu_2\sum_{j=1}^{M_i}\beta_{j}+\gamma_1\left(\sum_{j=1}^{M_i} \alpha_{j}h_{ij} - 1\right) + \gamma_2\sum_{j=1}^{M_i} \beta_{j}h_{ij} + \gamma_3\sum_{j=1}^{M_i} \alpha_{j}k_{ij} + \gamma_4\left(\sum_{j=1}^{M_i} \beta_{j}k_{ij} - 1\right)
\end{equation}
where $\Phi$ is an objective function as follows,
\begin{equation}
    \Phi_i = \sum_{j=1}^{M_i}w_{j}\left[(f_j - f_i) - h_{ij}\sum_{k=1}^{M_i}\alpha_{k}(f_k - f_i) - k_{ij}\sum_{k=1}^{M_i}\beta_{k}(f_k - f_i) \right]^2
\end{equation}
The minimization of $\Phi_i$ subject to the constraints, is given by $\nabla L = 0$. This yields
\begin{equation}\label{eq:system_PQ}
    PX = Q
\end{equation}
where 
\begin{equation*}
  P=\left[\begin{array}{cc}
R & S \\
S^T & O
\end{array}\right]  
\end{equation*}

\begin{equation*}
\mathrm{R}=\left[\begin{array}{ccc}
A & & 0 \\
& \ddots & \\
0 & & A
\end{array}\right]_{2M_i \times 2M_i}
\end{equation*}

\begin{equation*}
S=\left[\begin{array}{c}
S_1 \\
\vdots \\
S_{M_i}
\end{array}\right]_{2M_i \times 6}
\end{equation*}

\begin{equation*}
S_{j}=\left[\begin{array}{cccccc}
1 & 0  & h_{ij} & 0  & k_{ij} & 0 \\
0 & 1  & 0 & h_{ij}  & 0 & k_{ij}  \\
\end{array}\right]
\end{equation*}

\begin{equation*}
X^T=\left[\alpha_{1}, \beta_{1}, \ldots, \alpha_{M_i}, \beta_{M_i}, \mu_1, \mu_2, \gamma_1, \gamma_2, \gamma_3, \gamma_4\right]
\end{equation*}

\begin{equation*}
Q^T= \left[w_{1} h_{i1}, w_{1} k_{i1}, \ldots, w_{M_i} h_{iM_i}, w_{M_i} k_{iM_i}, 0,0,1,0,0,1 \right]
\end{equation*}
The meshless coefficients $\alpha_j, \beta_j$ can be extracted from the first $2M_i$ components of the solution vector $X$. The spatial derivatives of $f$ at the center point $i$ of cloud $C_i$ can also be evaluated using the midpoint $ij$ between $i$ and $j$ as
\begin{equation} \label{GCspatialderivative}
    a_1 = \left.\frac{\partial f}{\partial x}\right|_i = \sum_{j=1}^{M_i} \alpha_{ij}(f_{ij} - f_i) \mbox{~}, \mbox{ and } 
    a_2 = \left.\frac{\partial f}{\partial y}\right|_i = \sum_{j=1}^{M_i} \beta_{ij}(f_{ij} - f_i)
\end{equation}
where $\alpha_{ij}$ and $\beta_{ij}$ are the midpoint meshless coefficients, which are twice those evaluated using the satellite points $j$. The $f_{ij}$ is then estimated at the midpoint between $i$ and $j$ using the HLL Riemann solver as described in the subsequent sections.

\subsection{Spatial discretization}
The shallow water equations (Eq. \ref{goveqn}) can be expressed in terms of spatial derivatives within a cloud $C_i$ of points as
\begin{equation}
    \left.\frac{\partial \mathbf{U}}{\partial t}\right|_{C_{i}}+\nabla \cdot(\mathbf{G}, \mathbf{H})\bigg|_{C_{i}}=S_i
\end{equation}
Again, it can be expressed in the following compact form using the meshless coefficients obtained by solving \cref{eq:system_PQ} as
\begin{equation}\label{discretizedeq2}
    \frac{\partial \textbf{U}_i}{\partial t} + \sum_{j = 1}^{M_i} \lambda_{ij}(\textbf{F}_{ij} - \textbf{F}_i) = \textbf{S}_i
\end{equation}
where 
\begin{equation}
    \textbf{F} = \eta_x \textbf{G} + \eta_y \textbf{H}
\end{equation}
in which the vector $\vec{\eta} = (\eta_x, \eta_y)$ is estimated by $\displaystyle \eta_x = \frac{\alpha}{\lambda}, \eta_y = \frac{\beta}{\lambda}; \lambda = \sqrt{\alpha^2 + \beta^2}$. The flux function at the midpoint is computed by
\begin{equation} \label{mid_flux}
    \textbf{F}_{ij} = \textbf{F}(\textbf{U}_{ij}^L, \textbf{U}_{ij}^R)
\end{equation}

\begin{figure}[hbt!]
     \centering
     \subfloat[]{
     \begin{minipage}[b]{0.5\textwidth}
         \centering
         \includegraphics[width=0.7\textwidth]{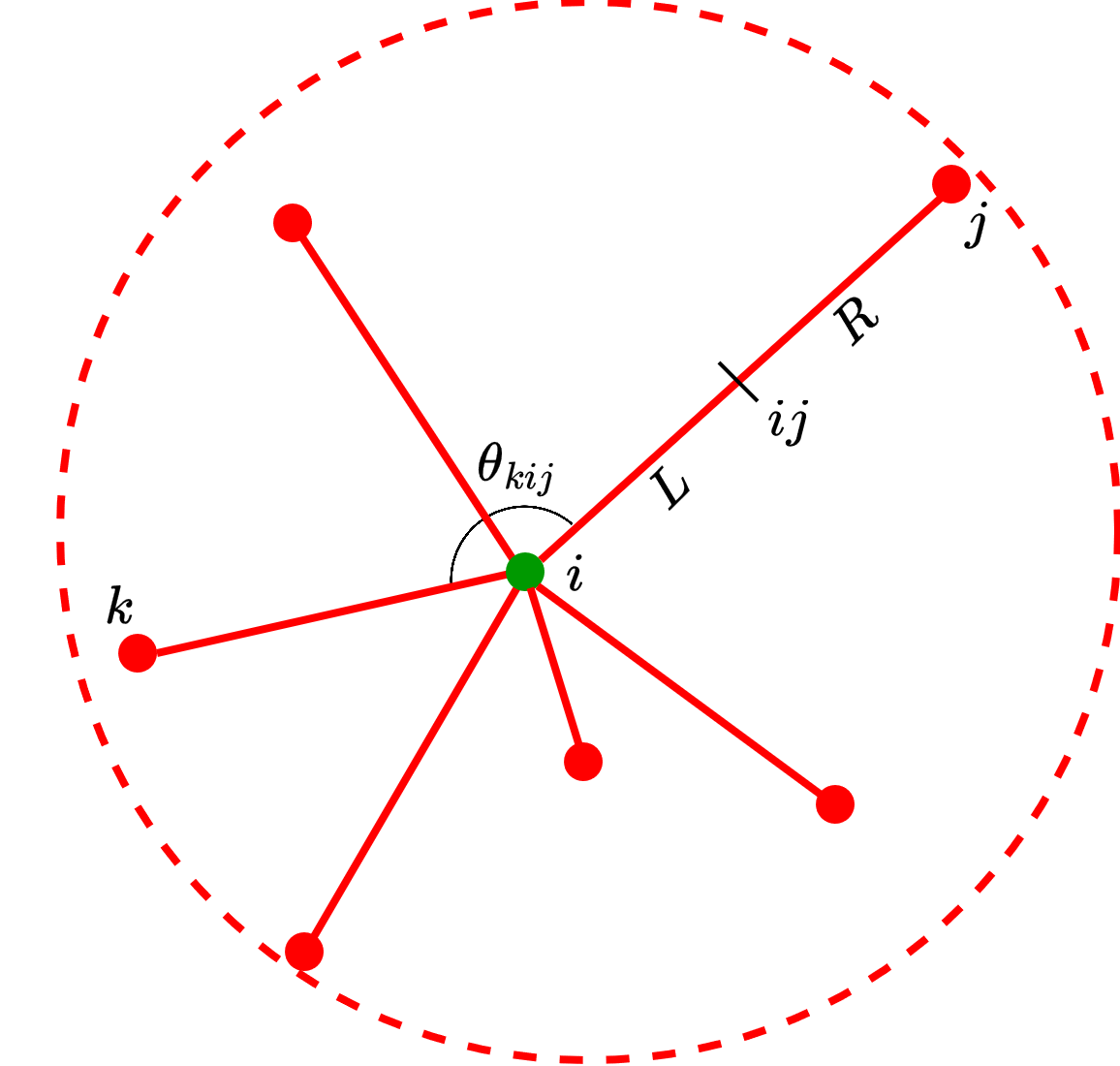}
     \end{minipage}}
     \subfloat[]{
     \begin{minipage}[b]{0.5\textwidth}
         \centering
         \includegraphics[width=0.8\textwidth]{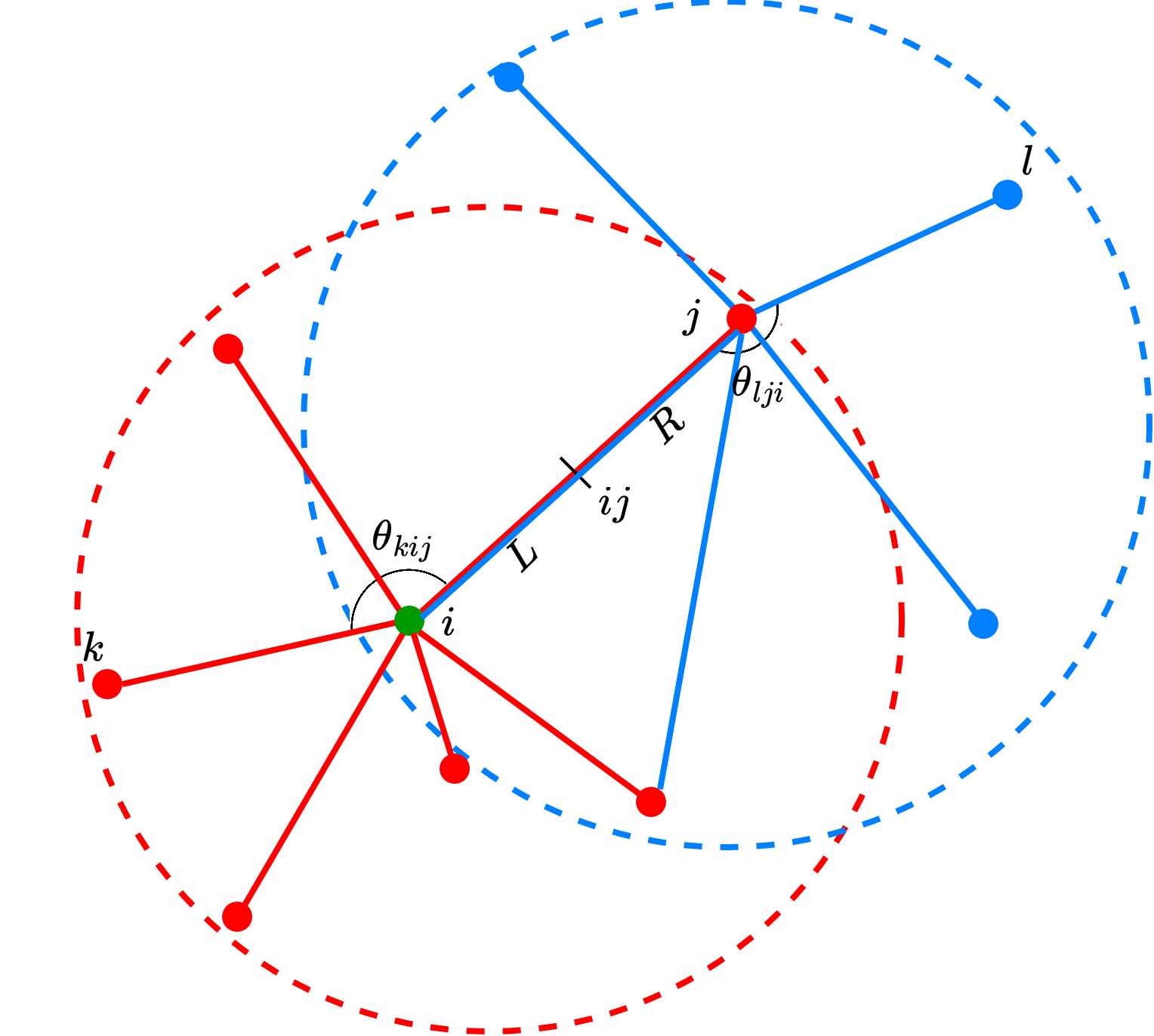}
     \end{minipage}}
        \caption{Minmod limiter for meshless method: (a) measuring angle $\theta_{kij}$ considering $i$ as center (b) measuring angle $\theta_{lji}$ considering $j$ as center}
        \label{fig:minmod_1}
\end{figure}

If $\boldsymbol{U}_{i j}^L=\boldsymbol{U_i}$ and $\boldsymbol{U}_{i j}^R=\boldsymbol{U_j}$ are chosen for computing the flux in \cref{mid_flux} through the midpoint, the numerical scheme results in a first-order accurate scheme, which is often diffusive in nature. For higher-order spatial accuracy, the variables are reconstructed at the midpoint $ij$ between the centre $i$ and its satellite $j$ given by  

\begin{equation}
    \textbf{U}_{ij}^L = \textbf{U}_i + \frac{\varphi_L}{2}(\textbf{U}_j - \textbf{U}_i)\\
    \textbf{U}_{ij}^R = \textbf{U}_j - \frac{\varphi_R}{2}(\textbf{U}_j - \textbf{U}_i)
\end{equation}

where $\varphi$ is a slope limiter \cite{Flux_limiters}. In the present work, the minmod slope limiter for meshfree method \cite{GCWLS} is employed to avoid under and over shooting of variables.
\begin{equation}
    \varphi_L =  max(0, min(1, r_{ij}^L))\\
    \varphi_R =  max(0, min(1, r_{ij}^R))
\end{equation}
\begin{equation}
    r_{ij}^L = \frac{s_{ki}}{s_{ji}} \cos{\theta_{kij}} \\
    r_{ij}^R = \frac{s_{lj}}{s_{ij}} \cos{\theta_{lji}}
\end{equation}
\begin{equation}
    s_{ki} = \displaystyle \frac{\textbf{U}_{k} - \textbf{U}_{i}}{||\boldsymbol{x}_{k} - \boldsymbol{x}_{i}||} 
\end{equation}
where $k$ is a satellite point in the cloud $C_i$ such that the angle $\theta_{kij}$ between two line segments $ki$ and $ij$ is the maximum as shown in \cref{fig:minmod_1}(a), and $l$ is a satellite point in the cloud $C_j$ such that the angle $\theta_{lji}$ between two line segments $lj$ and $ji$ is the maximum as shown in \cref{fig:minmod_1}(b).  


The interface fluxes $\mathbf{F}_{ij}$ (\cref{mid_flux}) are evaluated by solving the local Riemann problems defined at each neighbourhood. For this purpose, we have implemented the HLL approximate Riemann solver proposed by Harten, Lax and van Leer \cite{HLL}. The HLL Riemann solver computes the numerical flux in each neighbourhood, as described in the following section.

\subsection{HLL approximate Riemann solver}
Due to its robustness and straightforward implementation, the HLL approximate Riemann solver is used to compute convective fluxes at the midpoint $\mathbf{F}_{ij}$ (\cref{mid_flux}). This solver is particularly advantageous because it effectively handles discontinuities and provides stable solutions across various flow conditions. The HLL solver works by approximating the solution of the Riemann problem with three averaged states. These states represent the fluxes across different wave speeds, simplifying the complex interaction between waves emanating from discontinuities at the interfaces. By using averaged states, the HLL solver efficiently captures the essential features of the flow, such as shock waves and rarefactions. The approximate solution can be represented using three averaged states as follows
\begin{equation}
    \mathbf{U}_{ij} = \left\{ \begin{array}{ll}
    \mathbf{U}_L & \mbox{if } S_L \geq 0\\ 
    \mathbf{U}^* & \mbox{if } S_L < 0 < S_R\\ 
    \mathbf{U}_R & \mbox{if } S_R \leq 0\\ 
    \end{array}\right.
\end{equation}
The corresponding midpoint flux is given by
\begin{equation}\label{HLL_mid_flux}
    \mathbf{F}_{ij} \cdot \mathbf{n} = \left\{ \begin{array}{ll}
    \mathbf{F}_L \cdot \mathbf{n} & \mbox{if } S_L \geq 0\\ 
    \mathbf{F}^* & \mbox{if } S_L < 0 < S_R\\ 
    \mathbf{F}_R \cdot \mathbf{n} & \mbox{if } S_R \leq 0\\ 
    \end{array}\right.
\end{equation}
where $\mathbf{F_L} = \mathbf{F(U_L)}$ and $\mathbf{U_L}$ are the flux and conservative variable vectors, respectively evaluated at the left-hand side of the center of each cloud, while $\mathbf{F_R} = \mathbf{F(U_R)}$ and $\mathbf{U_R}$ are the same quantities evaluated at the right-hand side. $\mathbf{F^*}$ denotes the flux at the intermediate state, given by
\begin{equation}
    \mathbf{F}^{*}=\frac{S_{R} \mathbf{F}_{L} \cdot \mathbf{n} -S_{L} \mathbf{F}_{R} \cdot \mathbf{n} +S_{L} S_{R}\left(\mathbf{U}_{R}-\mathbf{U}_{L}\right)}{S_{R}-S_{L}}
\end{equation}
The symbols $S_L$ and $S_R$ represent the celerity of the waves, separating the constant states of the local Riemann problem. In the case of both sides are wet, the wave speeds $S_L$ and $S_R$ are given by
\begin{subequations}
\begin{align}
    S_{L} &= \min \left(\mathbf{q}_{L} \cdot \mathbf{n} -\sqrt{gh_{L}}, u^{*}-\sqrt{gh^{*}}\right)\\
    S_{R} &= \max \left(\mathbf{q}_{R} \cdot \mathbf{n} +\sqrt{gh_{R}}, u^{*}+\sqrt{gh^{*}}\right)
\end{align}
\end{subequations}
where
\begin{subequations}
\begin{align}
    u^{*} &= \frac{1}{2}\left(\mathbf{q}_{L}+\mathbf{q}_{R}\right) \cdot \mathbf{n} +\sqrt{gh_{L}}-\sqrt{gh_{R}}\\
    \sqrt{gh^{*}} &=\frac{1}{2}\left(\sqrt{gh_{L}}+\sqrt{gh_{R}}\right)+\frac{1}{4}\left(\mathbf{q}_{L}-\mathbf{q}_{R}\right) \cdot \mathbf{n}\\
   \mbox{ and ~~~} \mathbf{q} &= (u, v)^T
\end{align}
\end{subequations}
Note that for a wet-dry bed problem, the wave speeds $S_L$ and $S_R$ are estimated according to the following expressions
\begin{equation}
    S_{L}=\mathbf{q}_{L} \cdot \mathbf{n}-\sqrt{gh_{L}}, \quad S_{R}=\mathbf{q}_{L} \cdot \mathbf{n}+2 \sqrt{gh_{L}} \mbox{~~}\text { for right dry bed }
\end{equation}
\begin{equation}
    S_{L}=\mathbf{q}_{R}\cdot \mathbf{n} -2 \sqrt{gh_{R}}, \quad S_{R}=\mathbf{q}_{R}\cdot \mathbf{n} +\sqrt{gh_{R}} \mbox{~~}\text { for left dry bed }
\end{equation}
where $\mathbf{q}_L$ and $\mathbf{q}_R$ are velocity vectors of the left and right states, respectively; and $h_L$ and $h_R$ are water depths of the left and right states.

\subsection{Source term treatment}
The bottom topography of flow in a real-world situation may be highly irregular. Numerical treatment for the water surface gradient should also be properly estimated to maintain a stable and well-balanced scheme. The source term of water surface gradient has the following form
\begin{equation}
\mathbf{S}_{b}=\begin{bmatrix}
0 \\
-gh\displaystyle\frac{\strut\partial Z}{\strut\partial x}\\
-gh\displaystyle\frac{\strut\partial Z}{\strut\partial y}
\end{bmatrix}
\end{equation}
We estimated the water surface gradient for the second-order scheme with meshless coefficients, as expressed by 
\begin{equation} \label{surface_gradient1}
    \left(\frac{\partial Z}{\partial x}\right)_{C_i} = \sum_{j = 1}^{M_i} \alpha_{ij} \left(\overbar{Z}_{ij} - Z_i \right)
\end{equation}
\begin{equation} \label{surface_gradient2}
    \left(\frac{\partial Z}{\partial y}\right)_{C_i} = \sum_{j = 1}^{M_i} \beta_{ij} \left(\overbar{Z}_{ij} - Z_i \right)
\end{equation}
where
\begin{equation}
    \overbar{Z} = (Z_L + Z_R)/2
\end{equation}
The friction source term in \cref{sourceTerm} can be expressed as 
\begin{equation*}
\mathbf{S}_f = \begin{bmatrix}
0 \\
S_{fx} \\
S_{fy} \\
\end{bmatrix} = \begin{bmatrix}
0 \\
-C_f u \sqrt{u^2+v^2}\\
-C_f v \sqrt{u^2+v^2}\\
\end{bmatrix}
\end{equation*}
The friction source term is evaluated using a splitting point implicit scheme \cite{Liang, Bussing}, which is equivalent to solving the following ordinary differential equation

\begin{equation} \label{frictionDE}
    \frac{d\mathbf{U}}{dt} = \mathbf{S}_f
\end{equation}
Here, $\mathbf{U} = (hu, hv)^T = (q_x, q_y)^T$ of the momentum equations. Discretization of \cref{frictionDE} with implicit method gives
\begin{equation} \label{frictionImplicit}
    \frac{\mathbf{U}^{n+1} - \mathbf{U}^n}{\Delta t} = \mathbf{S}_f^{n+1}
\end{equation}
The non-zero components in friction term can be expanded using the first-order Taylor series as 
\begin{equation} \label{FrictionTaylor}
    S_{fx}^{n+1} = S_{fx}^n + \left(\frac{\partial S_{fx}}{\partial q_x}\right)^n \Delta q_x ,\\
    S_{fy}^{n+1} = S_{fy}^n + \left(\frac{\partial S_{fy}}{\partial q_y}\right)^n \Delta q_y
\end{equation}
where, $\Delta q = q^{n+1} - q^n$. Substituting \cref{FrictionTaylor} into \cref{frictionImplicit} and rearranging gives 
\begin{equation}
    q_x^{n+1} = q_x^n + \Delta t\left(\frac{S_{fx}}{D_x}\right)^n = q_x + \Delta t F_x , \\
    q_y^{n+1} = q_y^n + \Delta t\left(\frac{S_{fy}}{D_y}\right)^n = q_y + \Delta t F_y 
\end{equation}
where, $F_x and F_y$ are the implicit friction source term components. $\mathbf{D}$ is the implicit coefficient vector and
\begin{equation}
    \mathbf{D} = [D_x, D_y]^T = \left[1- \Delta t\left(\frac{\partial S_{fx}}{\partial q_x}\right)^n, 1 - \Delta t\left(\frac{\partial S_{fy}}{\partial q_y}\right)^n \right]^T.
\end{equation}

To prevent unrealistic friction arising from high fluid velocities, Liang and Marche \cite{Liang} implemented a limiting criteria for the friction force. This ensures that the primary impact of friction is to resist the fluid flow. Consequently, it indicates that the friction term cannot invert the direction of velocity components. The limiting value for $F_x$ and $F_y$ by considering
\begin{equation*}
    q_x^{n+1}q_x^n \geq 0, q_y^{n+1}q_y^n \geq 0
\end{equation*}
is 
\begin{equation}
    F_x \leftarrow \left\{\begin{tabular}{c c}
         $\displaystyle max(-q_x^n/\Delta t, F_x)$ & \text{if } $q_x^n \geq 0$,\\
         $\displaystyle min(-q_x^n/\Delta t, F_x)$ & \text{if } $q_x^n < 0$
    \end{tabular} \right.
    F_y \leftarrow \left\{\begin{tabular}{c c}
         $\displaystyle max(-q_y^n/\Delta t, F_y)$ & \text{if } $q_y^n \geq 0$,\\
         $\displaystyle min(-q_y^n/\Delta t, F_y)$ & \text{if } $q_y^n < 0$
    \end{tabular} \right.
\end{equation}
As explained above, limiting criteria to the friction terms of the momentum equations ensures enhanced model stability, especially in conditions where flow depth becomes small and velocity tends to be unrealistically very high.

\subsection{Wetting and drying treatment}
 The shallow water equations are solved in this study to simulate the propagation of the waterfront over a natural topography. The wetting and drying process refers to the phenomenon where areas of the computational domain transition between being submerged (wet) and exposed (dry) as the waterfront travels over the topography over time. This process is particularly important in simulating scenarios such as dam-break flow, flood wave propagation, levee breaching flow, etc. When computational nodes become dry, the shallow water equations are modified to prevent numerical instabilities that may arise from attempting to solve equations in dry areas. This often involves applying special dry-bed stabilization techniques (i.e., wetting and drying treatment) to ensure numerical stability.

\begin{figure}[hbt!]
     \centering
     \subfloat[]{
     \begin{minipage}[b]{0.5\textwidth}
         \centering
         \includegraphics[width=0.95\textwidth]{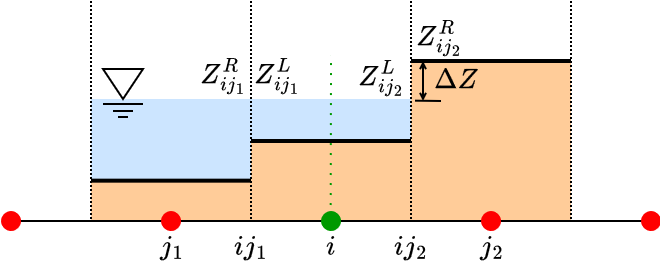}
     \end{minipage}}
     \subfloat[]{
     \begin{minipage}[b]{0.5\textwidth}
         \centering
         \includegraphics[width=0.9\textwidth]{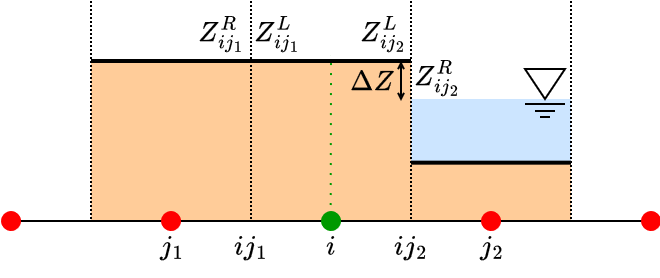}
     \end{minipage}}
        \caption{Local modification of water surface level: (a) $h_L \ne 0, h_R = 0$ and $Z_L < Z_R$(b) $h_L = 0, h_R \ne 0$ and $Z_R < Z_L$}
        \label{fig:dry_wet}
\end{figure}

 In this meshless method, the surface elevation $Z$ and bed elevation $z_b$ (\cref{fig:dry_wet}) are considered constant throughout the virtual cell of a point $i$ surrounded by its neighbours $j$ points. As we have solely integrated surface elevation into the discretization process, it suffices to locally adjust the surface elevation, $Z = h + z_b$, which automatically modifies the bed elevation. When a wet virtual cell is adjacent to a dry virtual cell, two cases arise (\cref{fig:dry_wet}), as shown in \cref{fig:dry_wet}: case (a)  $h_L \ne 0, h_R = 0$ and $Z_L < Z_R$, and case (b)  $h_L = 0, h_R \ne 0$ and $Z_R < Z_L$. In both cases, there should not be any flux through the midpoint of the interface $ij_2$. However, in our model, $Z^R_{ij_2}$ and $Z^L_{ij_2}$ as shown in \cref{fig:dry_wet} acts as water surfaces and creates a non-physical flow through the midpoint of the interface $ij_2$. To avoid this, we locally modify the water surface elevation by subtracting $\Delta Z = \left|Z^R_{ij_2} - Z^L_{ij_2}\right|$. 

The proposed wetting and drying treatment discussed above works without any issues for the proposed meshless method.

\subsection{Temporal discretization}
The governing equations, represented by \cref{goveqn}, are converted into a spatially discretized form as in \cref{discretizedeq2} to facilitate the numerical solution. This conversion involves breaking down the governing equations into discretization of the fluxes (\cref{HLL_mid_flux}), water surface gradient in the source terms (\cref{surface_gradient1} and \cref{surface_gradient2}), and friction terms (\cref{FrictionTaylor}), that can be computed at discrete points in space. Combining all these discretized forms, the governing equation (\cref{goveqn}) is now reduced to a simple temporal equation, as written below.
\begin{equation} \label{semidis}
    \frac{\partial \mathbf{U}}{\partial t} = \mathbf{R}
\end{equation}
where
\begin{equation*}
    \mathbf{R} = \mathbf{S}-\frac{\partial \mathbf{F}}{\partial x}
\end{equation*}
The \cref{semidis} can now be solved explicitly using the forward Euler time discretization for every cloud $C_i$, given below.
\begin{equation}
    \frac{\mathbf{U}_i^{n+1} - \mathbf{U}_i^n}{\Delta t} = \mathbf{R}_i
\end{equation}
where $n$ and $n+1$ denote the physical quantities at the $n^{th}$ and $(n+1)^{th}$ time steps, respectively. Therefore, the vector of the conservative variables at $(n+1)^{th}$ time step can be obtained by
\begin{equation*}
    \mathbf{U}_i^{n+1} =  \mathbf{U}_i^n + (\Delta t) \mathbf{R}_i
\end{equation*}




\subsection{Boundary conditions}
The proposed meshless method is implemented to work according to the flow regime at the boundaries. Additional points are introduced beyond the actual flow domain to apply suitable boundary conditions. The new points may be assumed to be ghost points just to apply boundary conditions. In cases of subcritical flow, one must designate a boundary condition at each inflow and outflow boundary. When inflow is subcritical, discharge is typically specified at the new point upstream, while the water surface level is determined by extrapolating adjacent values from interior points. Similarly, the water surface level is defined at a new point downstream, with discharge being extrapolated from adjacent values from interior points. Discharge and water level are defined at the new upstream point for instances of supercritical inflow. For transmissive boundaries (i.e., open outflow), the same conserved variables as those at neighbouring interior points within the domain are assigned to the new exterior points.

\section{Numerical tests}
The proposed meshless method within the framework of shock-capturing capability is verified by solving 1D and 2D analytical tests. The model is then validated against experimental observations. Finally, the model is applied to simulate a dam-break event in a large-scale physical model.

\subsection{Analytical 1D dam-break flow on wet and dry beds}
\begin{figure}[hbt!]
     \centering
     \subfloat[]{
     \begin{minipage}[b]{0.5\textwidth}
         \centering
         \includegraphics[width=\textwidth]{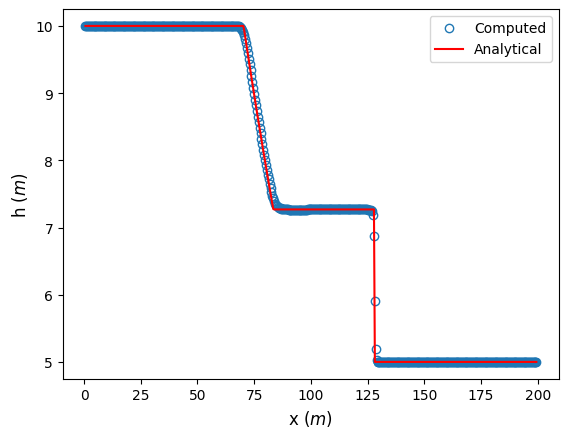}
     \end{minipage}}
     \subfloat[]{
     \begin{minipage}[b]{0.5\textwidth}
         \centering
         \includegraphics[width=\textwidth]{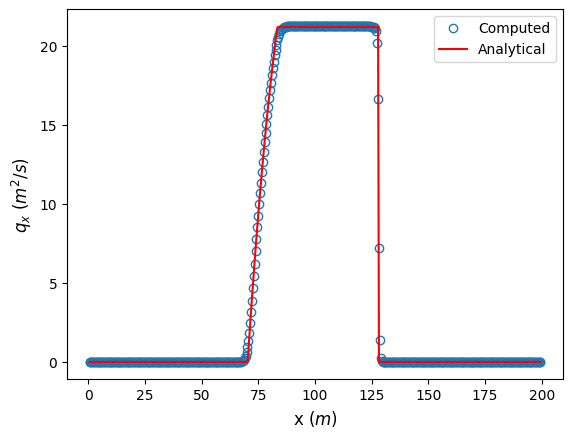}
     \end{minipage}}
        \caption{Dam-break on wet bed: (a) water surface elevation  (b) discharge}
        \label{fig:wetbed}
\end{figure}

\begin{figure}[hbt!]
     \centering
     \subfloat[]{
     \begin{minipage}[b]{0.5\textwidth}
         \centering
         \includegraphics[width=\textwidth]{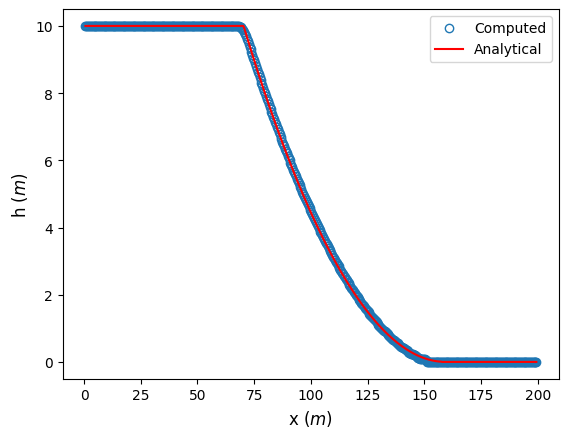}
     \end{minipage}}
     \subfloat[]{
     \begin{minipage}[b]{0.5\textwidth}
         \centering
         \includegraphics[width=\textwidth]{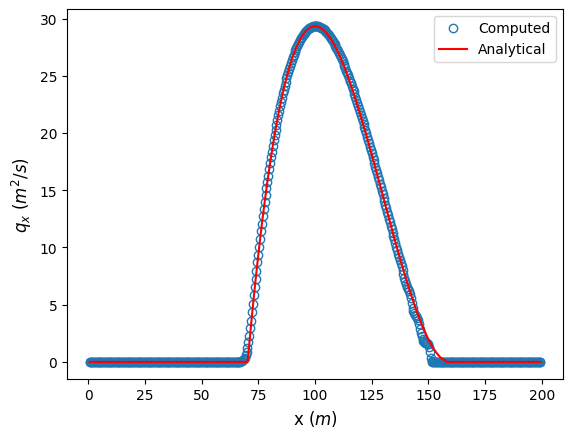}
     \end{minipage}} 
        \caption{Dam-break on dry bed: (a) water surface elevation  (b) discharge}
        \label{fig:drybed}
\end{figure}

This test case is conducted to assess the model's capability in simulating discontinuous flow, waterfront propagation, and its ability to capture shocks over wet and dry beds. A horizontal rectangular channel measuring 200 m in length and 10 m in width, featuring a frictionless bed, is considered. A dam is located at $x = 100 $ m from the upstream end. Initially, the upstream water depth is 10 m, while downstream water depths are 5 m for wet and 0 m for dry bed conditions. In the numerical computation, a depth tolerance of $h_{tol}=0.000001$ m is utilized to distinguish between wet and dry beds. The channel is described as having 15,020 irregularly distributed meshless points along its length. At $t=0$ s, the dam is instantaneously removed, generating a bore (i.e., shock) wave moving from left to right and a depression (i.e., rarefaction) wave propagating towards the left. Results are recorded at 3 s post-dam removal and compared against analytical solutions \cite{Toro, Henderson}. Flow depth and velocity for wet and dry bed scenarios are illustrated in \cref{fig:wetbed} and \cref{fig:drybed}, respectively. It is important to highlight that the proposed meshless method aptly captures sharp variations in depth and velocity profiles, thereby accurately simulating discontinuous flow with shocks over wet and dry beds.

\subsection{Moving shorelines in a 2D frictional parabolic bowl}
To verify the capability of the proposed method of accurately simulating flow with the wetting and drying process in a parabolic bed topography with friction. This benchmark test case also verifies the numerical treatment of bed friction and slope terms. The analytical solution of the moving shoreline in a 2D frictional parabolic bowl was developed by Sampson et al. \cite{Sampson2006}. Consider a rectangular computational domain [0 m, 8000 m] $\times$ [0 m, 8000 m] and bed topography with the centre $(x_0, y_0)$, described as 
\begin{equation}
    z_b(x, y)=h_0\left[\left(x-x_0\right)^2+\left(y-y_0\right)^2\right] / a^2
\end{equation}

where, $h_0$ and $a$ are constants. The analytical solution for the water level, for the case with bed frictional parameter $\tau$ is chosen less than the peak amplitude parameter $p$, is given by 

\begin{equation} \label{prabolic_initial_eq1}
    \begin{aligned}
\eta(x, y, t) & =h_0-\frac{1}{2 g} B^2 e^{-\tau t}-\frac{1}{g} B e^{-\tau t / 2} \\
& \times\left[\left(\frac{\tau}{2} \sin s t+s \cos s t\right) \times\left(x-x_0\right)+\left(\frac{\tau}{2} \cos s t-s \sin s t\right) \times\left(y-y_0\right)\right]
\end{aligned}
\end{equation}

and analytical solutions for the velocities are given by
\begin{equation} \label{prabolic_initial_eq2}
    u(t)=B e^{-\tau t / 2} \sin s t, \quad v(t)=B e^{-\tau t / 2} \cos s t,
\end{equation}

where $p=\sqrt{8gh_0}/a$; $B$ is a given constant; and $s = \sqrt{p^2 - \tau^2}/2$. The coefficient $C_f$ in bed friction term equals to $h\tau/\sqrt{u^2 + v^2}$.

\begin{figure}[hbt!]
     \centering
     \begin{minipage}[b]{0.90\textwidth}
         \centering
         \includegraphics[width=0.90\textwidth,valign=t]{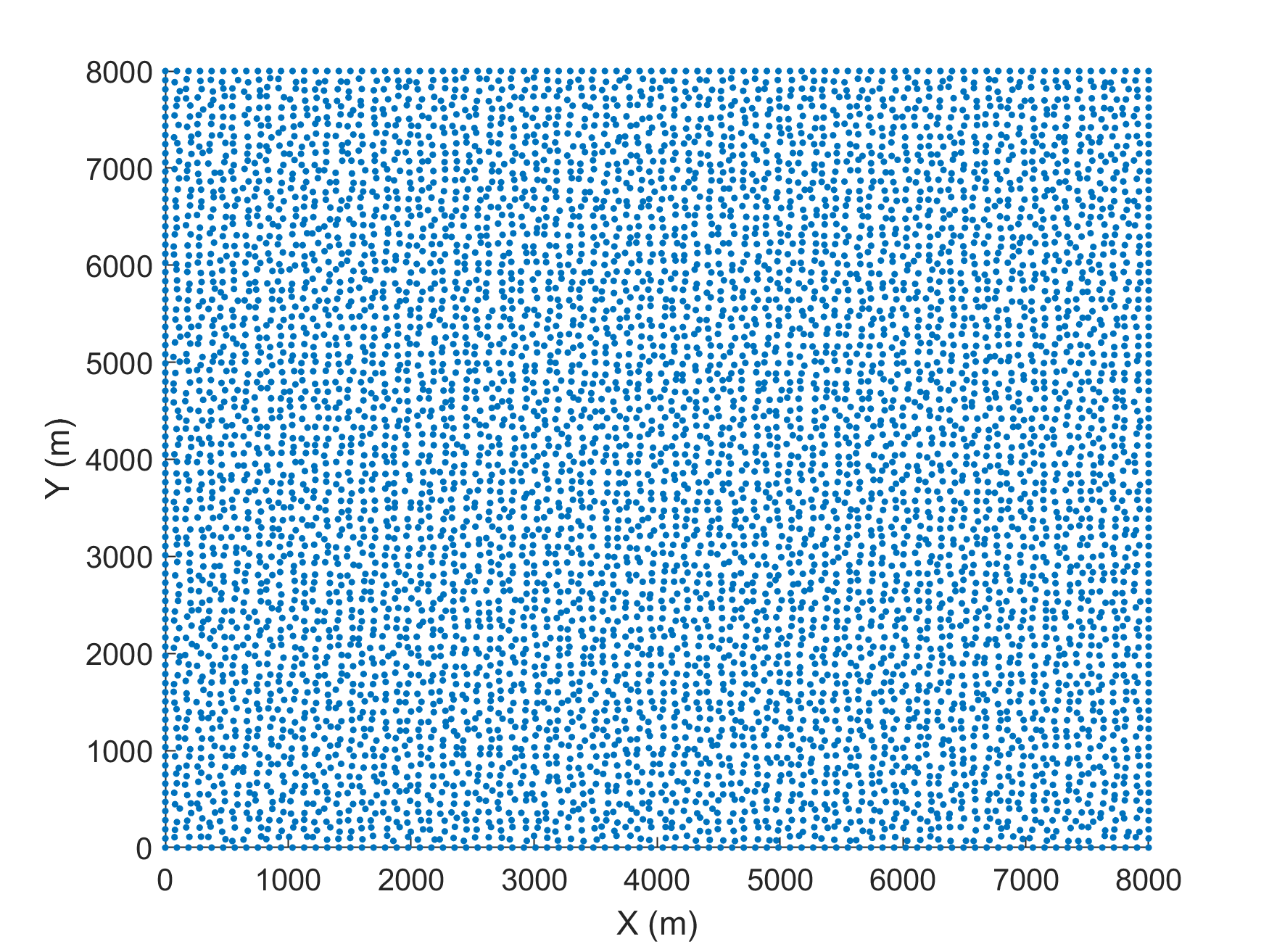}
     \end{minipage}
     \caption{Moving shorelines in a 2D frictional parabolic bowl: Computational domain and point distribution}
        \label{fig:parabolic_pointsDist}
\end{figure}

\begin{figure}[hbt!]
     \begin{minipage}[b]{0.90\textwidth}
         \centering
         \includegraphics[width=0.90\textwidth,valign=t]{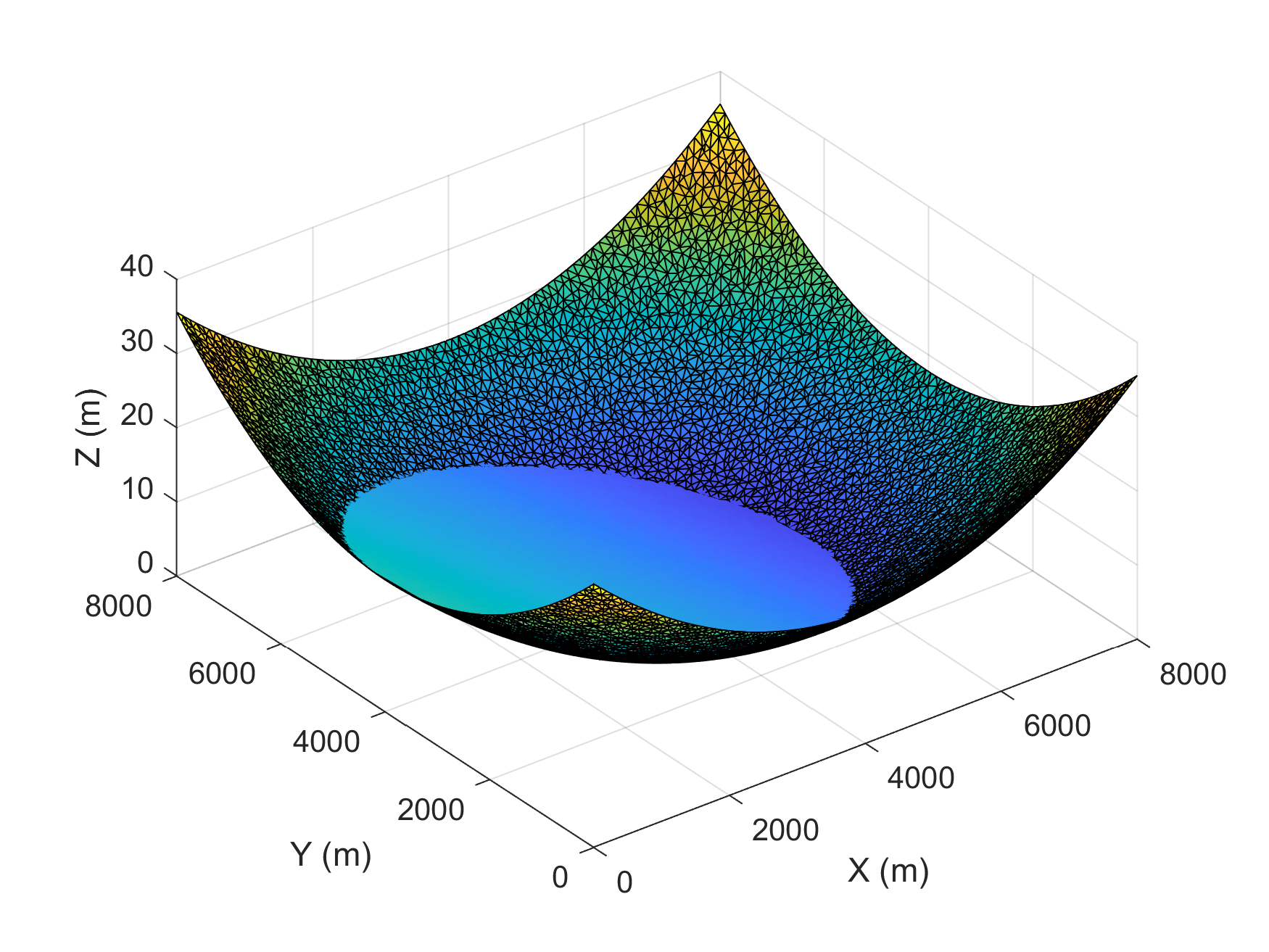}
     \end{minipage}
        \caption{Moving shorelines in a 2D frictional parabolic bowl: Initial water surface level}
        \label{fig:parabolic_points_initialwater}
\end{figure}

\begin{figure}[hbt!]
     \centering
     \subfloat[]{
     \begin{minipage}[b]{0.50\textwidth}
         \centering
         \includegraphics[width=\textwidth,valign=t]{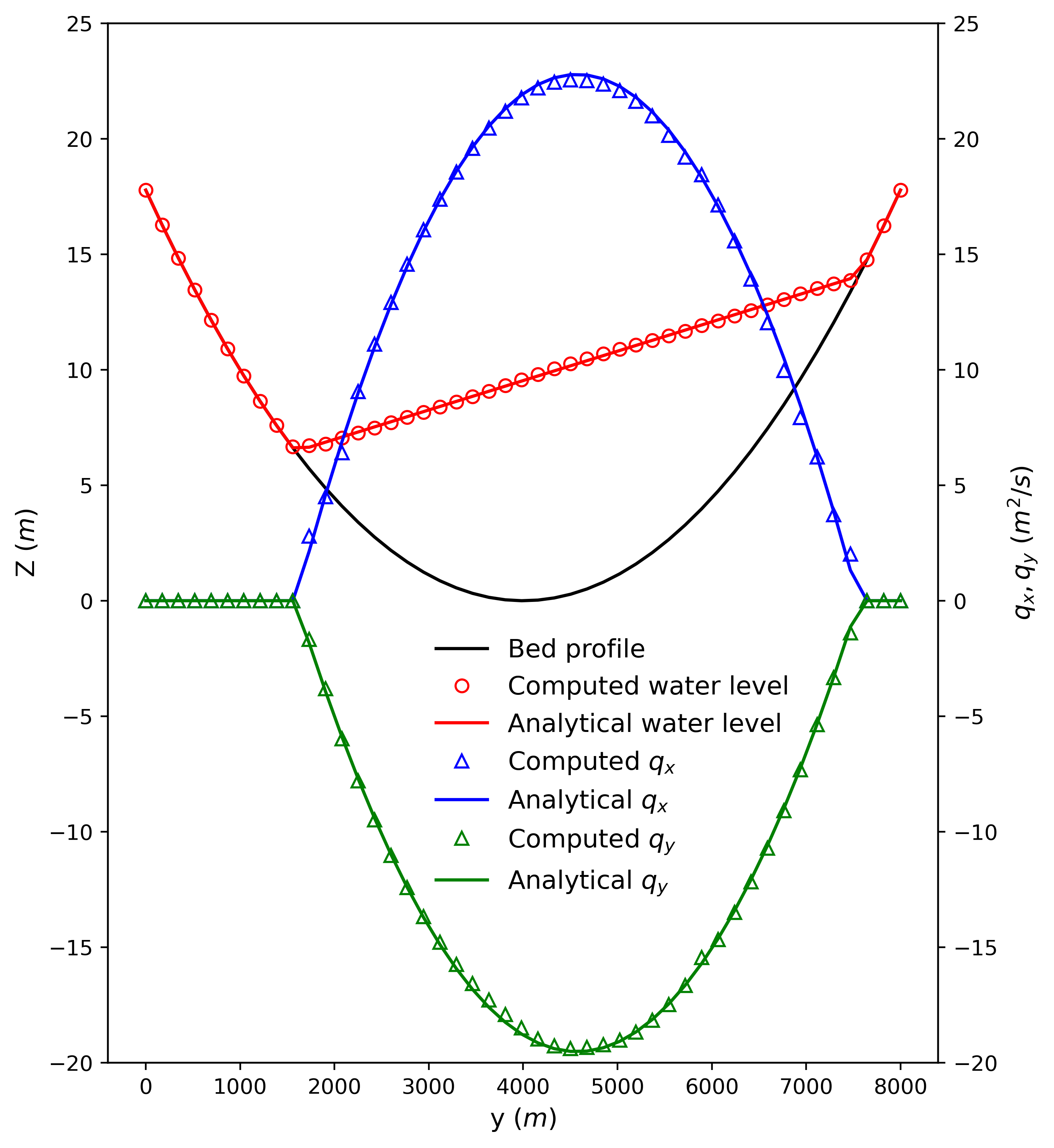}
     \end{minipage}}
     \subfloat[]{
     \begin{minipage}[b]{0.50\textwidth}
         \centering
         \includegraphics[width=\textwidth,valign=t]{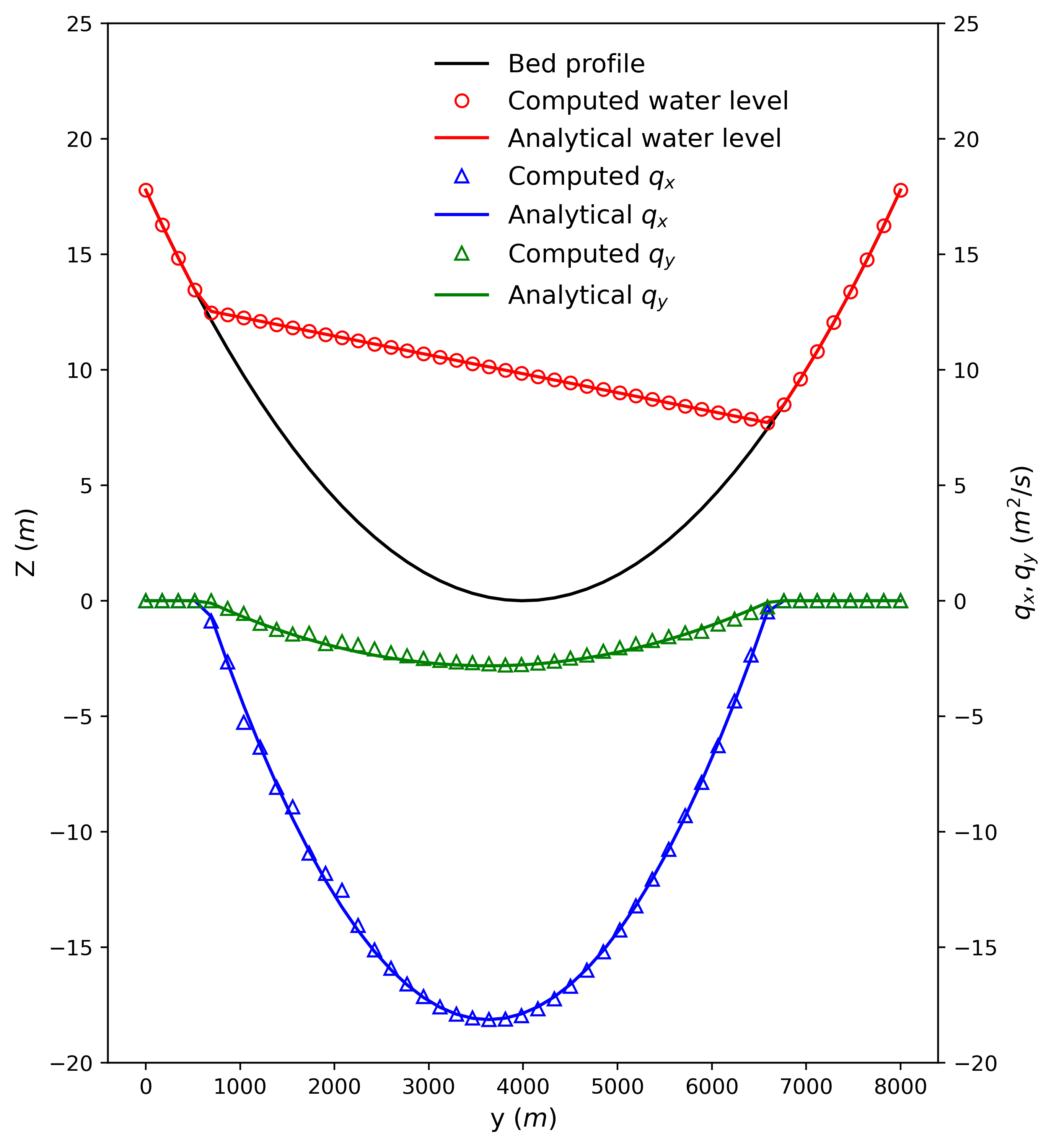}
     \end{minipage}} 
     \hfill
     \subfloat[]{
     \begin{minipage}[b]{0.50\textwidth}
         \centering
         \includegraphics[width=\textwidth,valign=t]{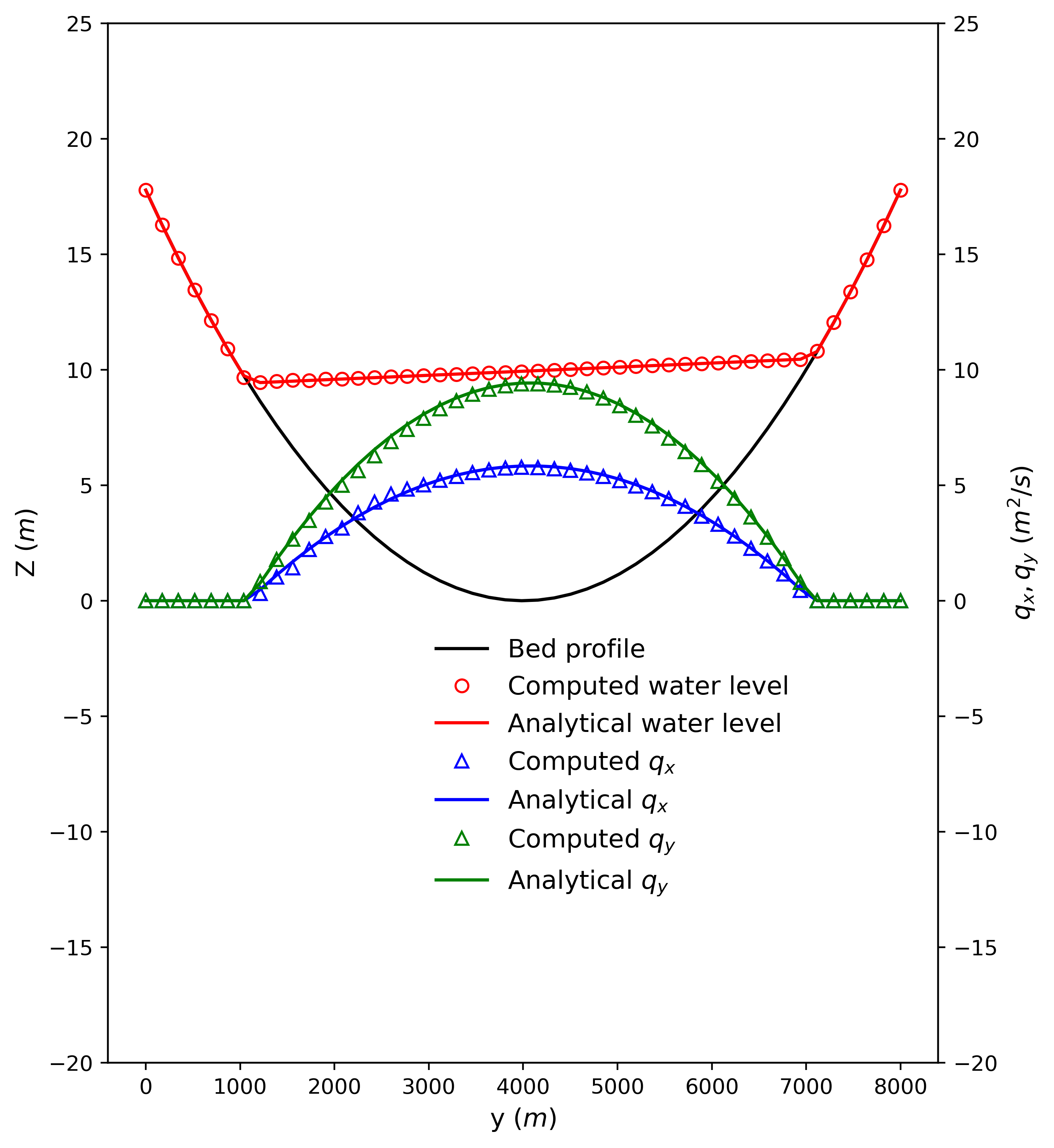}
     \end{minipage}}
     \subfloat[]{
     \begin{minipage}[b]{0.50\textwidth}
         \centering
         \includegraphics[width=\textwidth,valign=t]{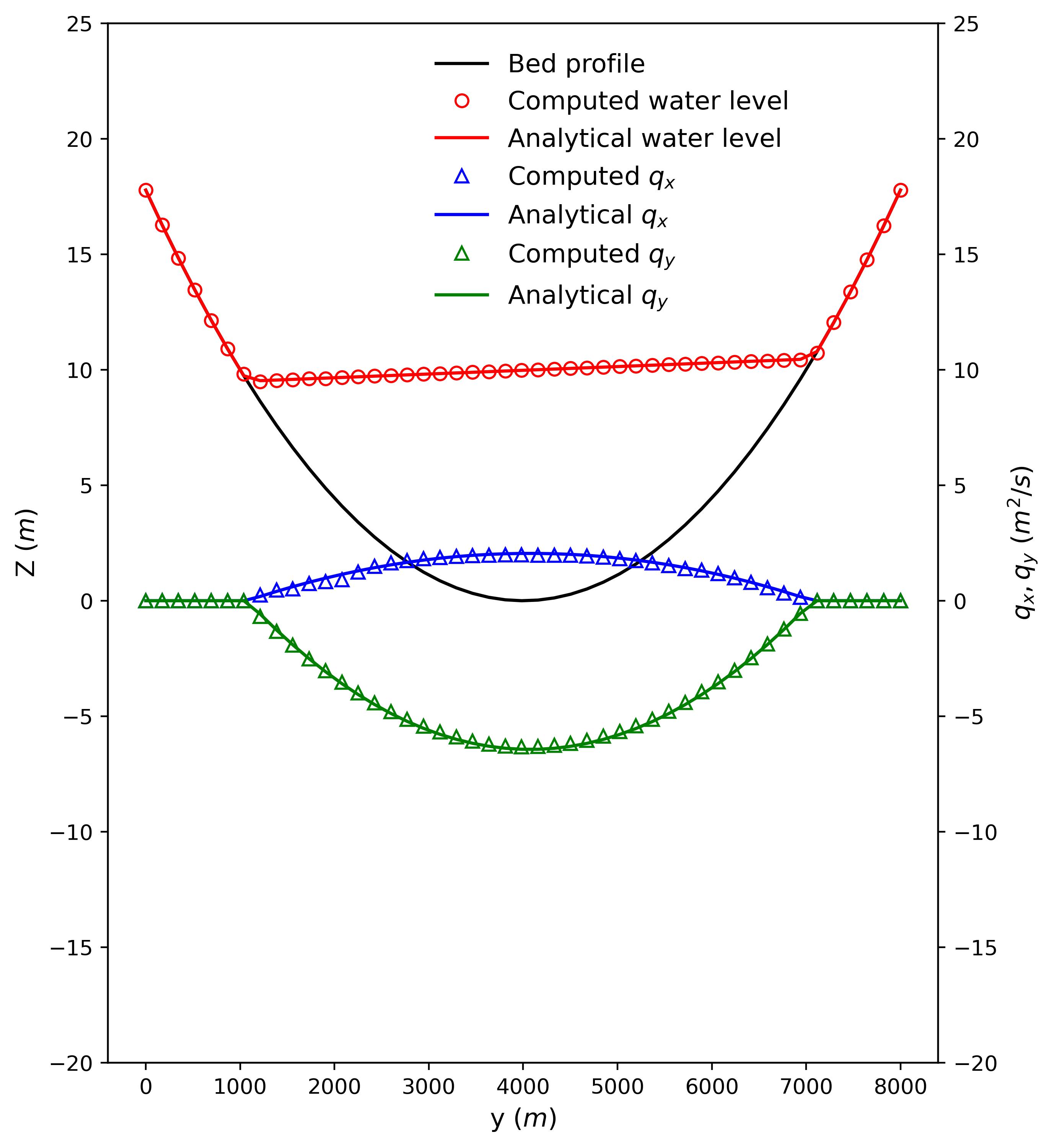}
     \end{minipage}}
        \caption{Moving shorelines in a 2D frictional parabolic bowl: (a) $t = 500 s$  (b) $t = 1000 s$  (c) $t = 1500 s$  (d) $t = 2000 s$}
        \label{fig:parabolic_results_at_different_times}
\end{figure}

\begin{figure}[hbt!]
     \centering
     \subfloat[]{
     \begin{minipage}[b]{0.50\textwidth}
         \centering
         \includegraphics[width=\textwidth,valign=t]{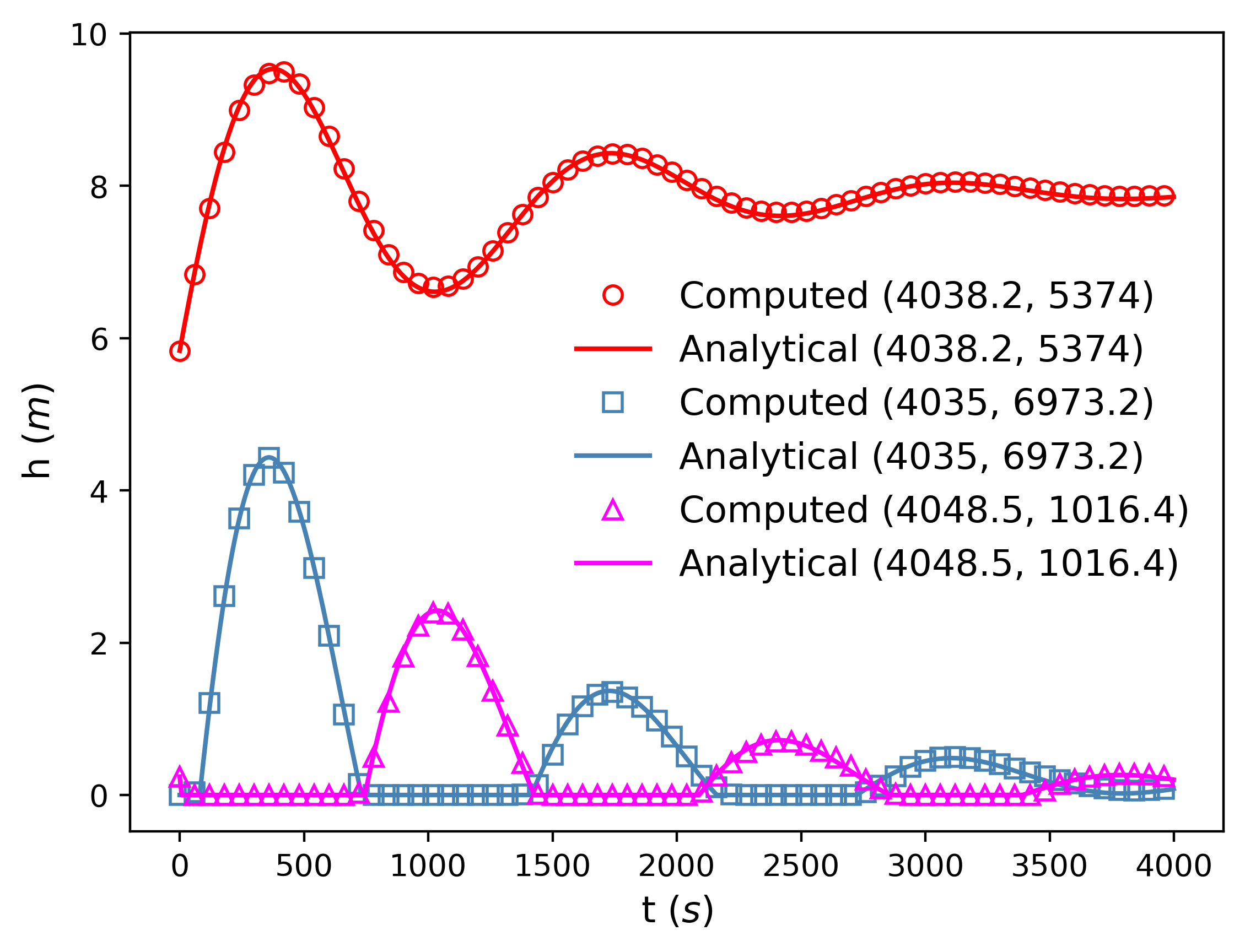}
     \end{minipage}}
     \subfloat[]{
     \begin{minipage}[b]{0.50\textwidth}
         \centering
         \includegraphics[width=\textwidth,valign=t]{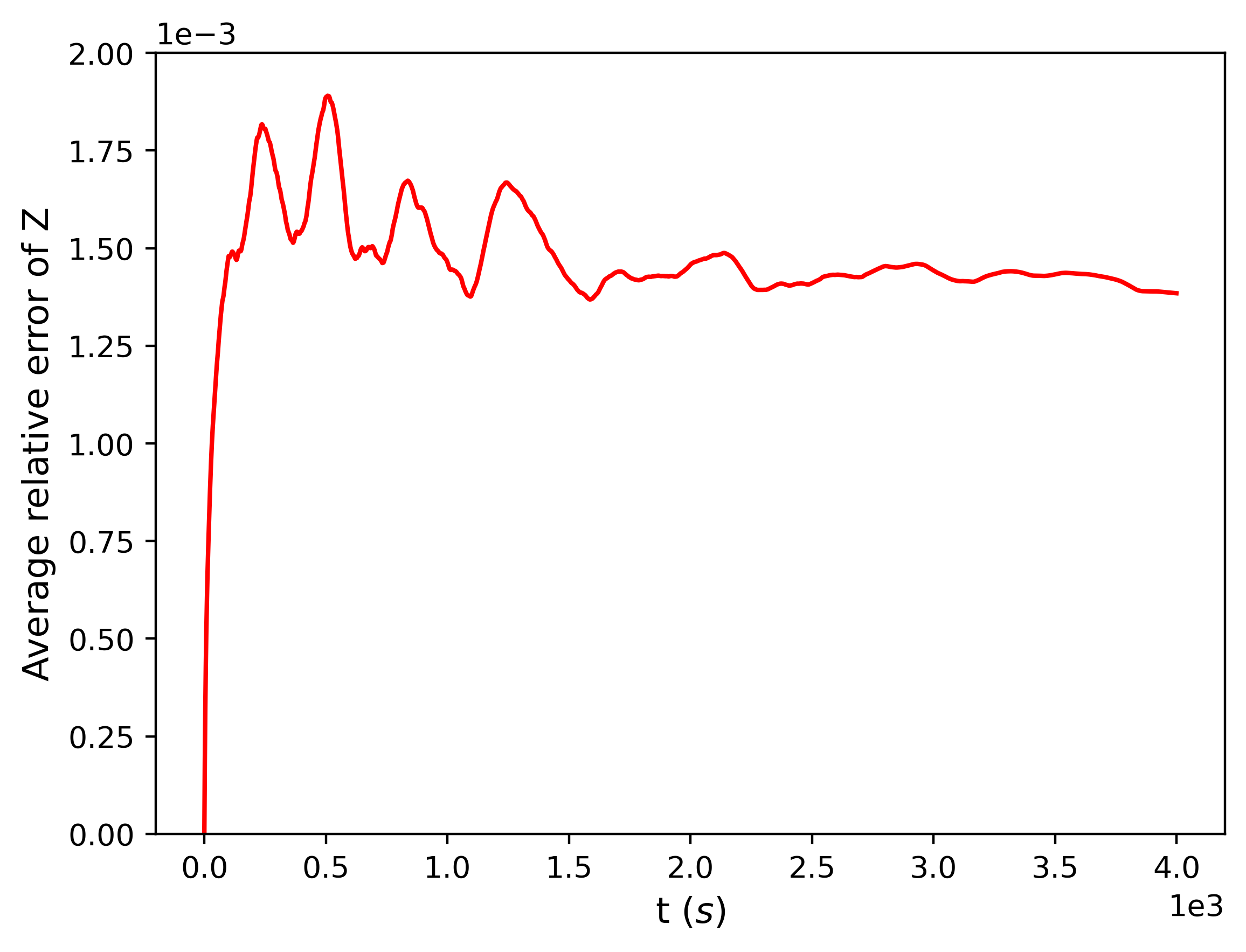}
     \end{minipage}} 
        \caption{Moving shorelines in a 2D frictional parabolic bowl: (a) Comparison of numerical and analytical results for water depths at three different points   (b) Average relative error of water surface level}
        \label{fig:parabolic_comparison}
\end{figure}

The computational domain is defined by filling 7396 meshless points as shown in \cref{fig:parabolic_pointsDist}. The parameters are set to $h_0 = 10 $ m, $a = 3000 $ m, $\tau = 0.002 s^{-1}$  and $B = 5 $ m/s. The initial conditions are given in \cref{prabolic_initial_eq1} and \cref{prabolic_initial_eq2} at $t= 0 $ s, and closed (i.e., wall) boundary condition is applied at all the four boundaries. We run the simulation until $t = 6000 $ s. \cref{fig:parabolic_results_at_different_times} shows the computed water level together with $q_x$ and $q_y$ along the line $x = 4000$ m at different times. The moving shoreline, $q_x$ and $q_y$ can be seen in excellent agreement with analytical solutions; thus, the moving shorelines with the wetting and drying procedure of the present method validate the discretization of the flux and source terms. \cref{fig:parabolic_comparison}(a) shows the comparison between numerical and analytical solutions for water depth at points (4038.2, 5374), (4035, 6973.2), and (4048.5, 1016.4). The average relative error is calculated for the water surface elevation to check the accuracy of the numerical solution and is evaluated by

$$L_1(Z) = \displaystyle \frac{1}{N}\sum_{i = 1}^{N}\left|\frac{Z_i - \tilde{Z_i}}{\tilde{Z_i}}\right|$$
where $Z_i$ and $\tilde{Z_i}$ are the numerical and analytical water surface elevation at the point $i$, respectively. \cref{fig:parabolic_points_initialwater} shows the average relative error of the water surface elevation.

\subsection{Dam break wave propagating over three humps}
This 2D dam-break flow over three humps has been widely considered a benchmark test case to deal with complicated wetting and drying process in realistic flood modelling applications. In this test case, a dam-break wave travels over an initially dry floodplain with three humps, which was originally proposed by Kawahara and Umestu \cite{Kawahara_Umetsu}. The dam-break occurs in a 75 m long and 30 m wide closed channel, with the bed topography defined by
\begin{equation}
    \begin{aligned}
z_{\mathrm{b}}(x, y)= & \max \left[0,1-\frac{1}{8} \sqrt{(x-30)^2+(y-6)^2},\right. 1-\frac{1}{8} \sqrt{(x-30)^2+(y-24)^2}, \\
& \left.3-\frac{3}{10} \sqrt{(x-47.5)^2+(y-15)^2}\right]
\end{aligned}
\end{equation}

\begin{figure}[hbt!]
     \centering
     \begin{minipage}[b]{0.90\textwidth}
         \centering
         \includegraphics[width=0.90\textwidth,valign=t]{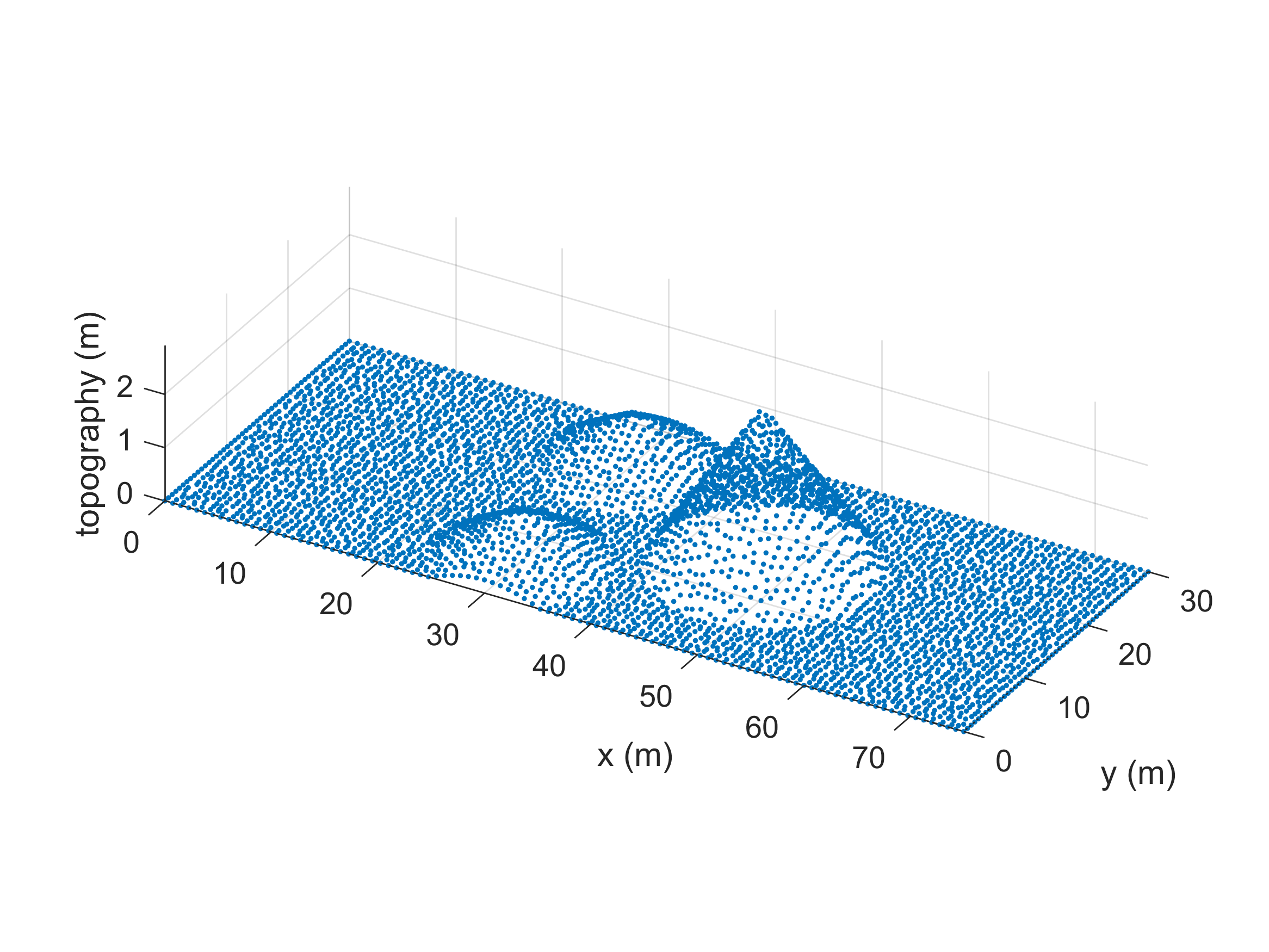}
     \end{minipage}
        \caption{Dam break wave propagating over three humps: (a) bottom topography and point distribution}
        \label{fig:ThreeHumps_Topography}
\end{figure}

\begin{figure}[hbt!]
     \centering
     \subfloat{
     \begin{minipage}[b]{0.5\textwidth}
         \centering
         \includegraphics[width=\textwidth]{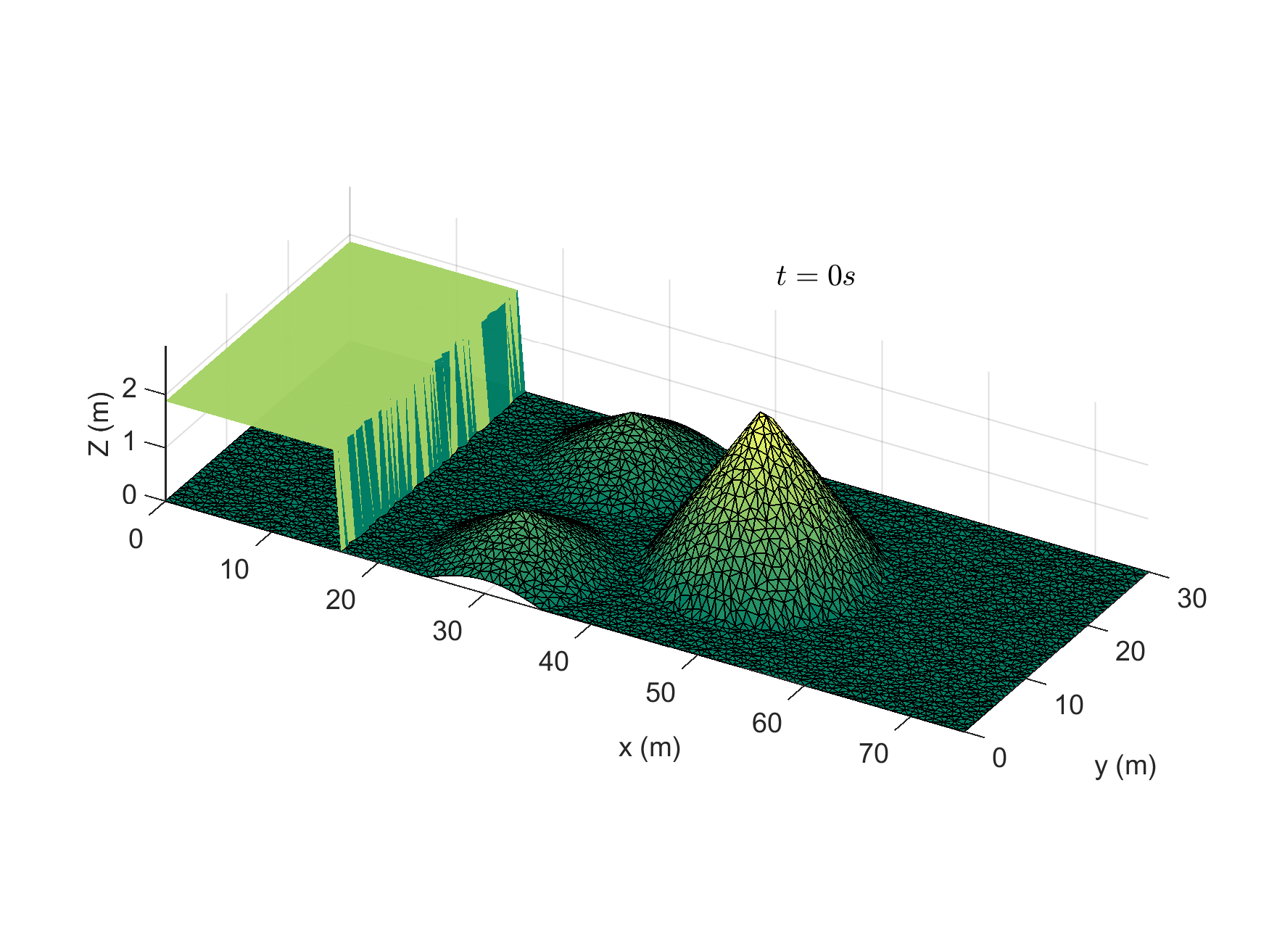}
     \end{minipage}}
     \subfloat{
     \begin{minipage}[b]{0.5\textwidth}
         \centering
         \includegraphics[width=\textwidth]{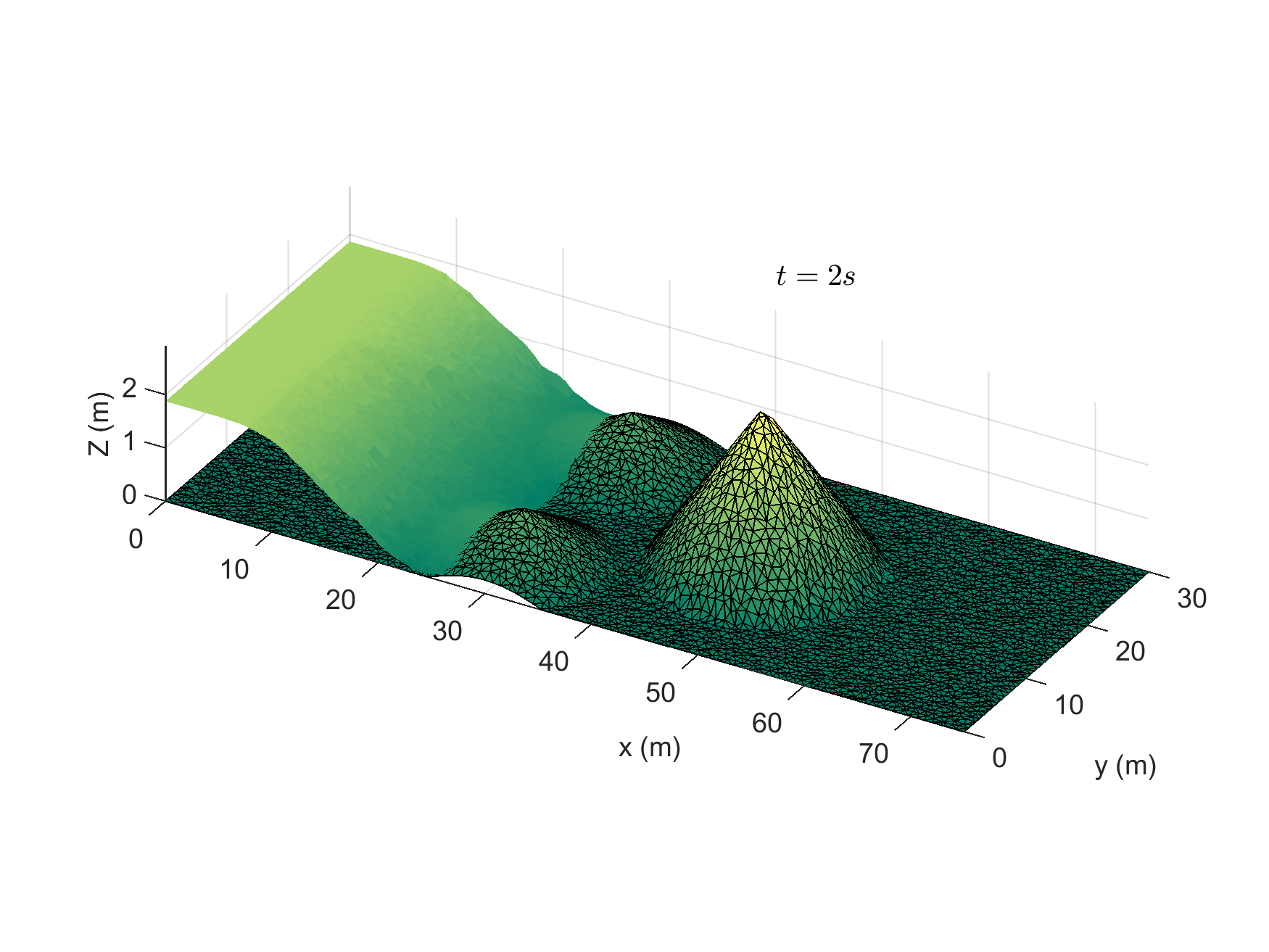}
     \end{minipage}}\\[-12ex]
     \hfill
     \subfloat{
     \begin{minipage}[b]{0.5\textwidth}
         \centering
         \includegraphics[width=\textwidth]{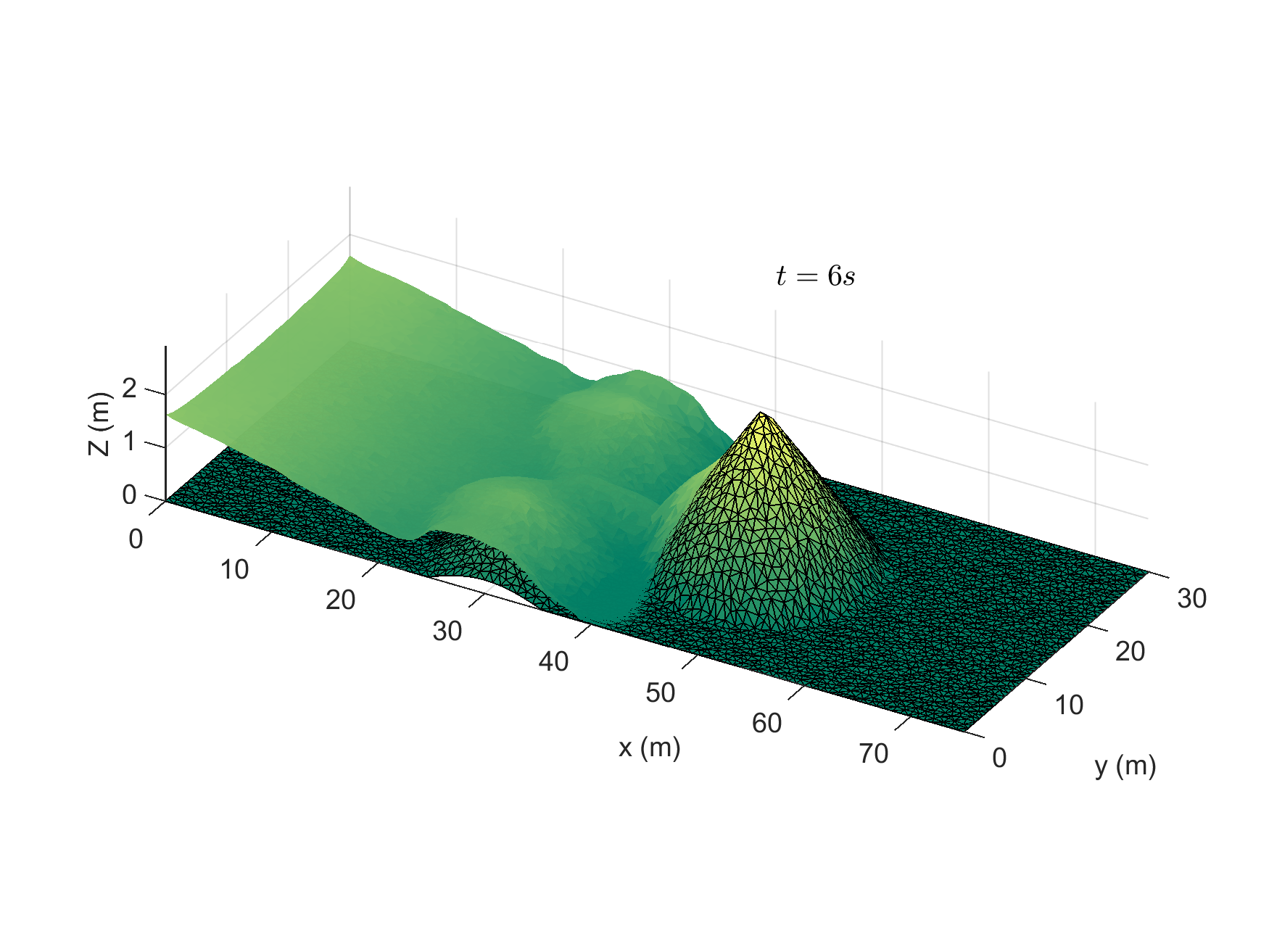}
     \end{minipage}}
     \subfloat{
     \begin{minipage}[b]{0.5\textwidth}
         \centering
         \includegraphics[width=\textwidth]{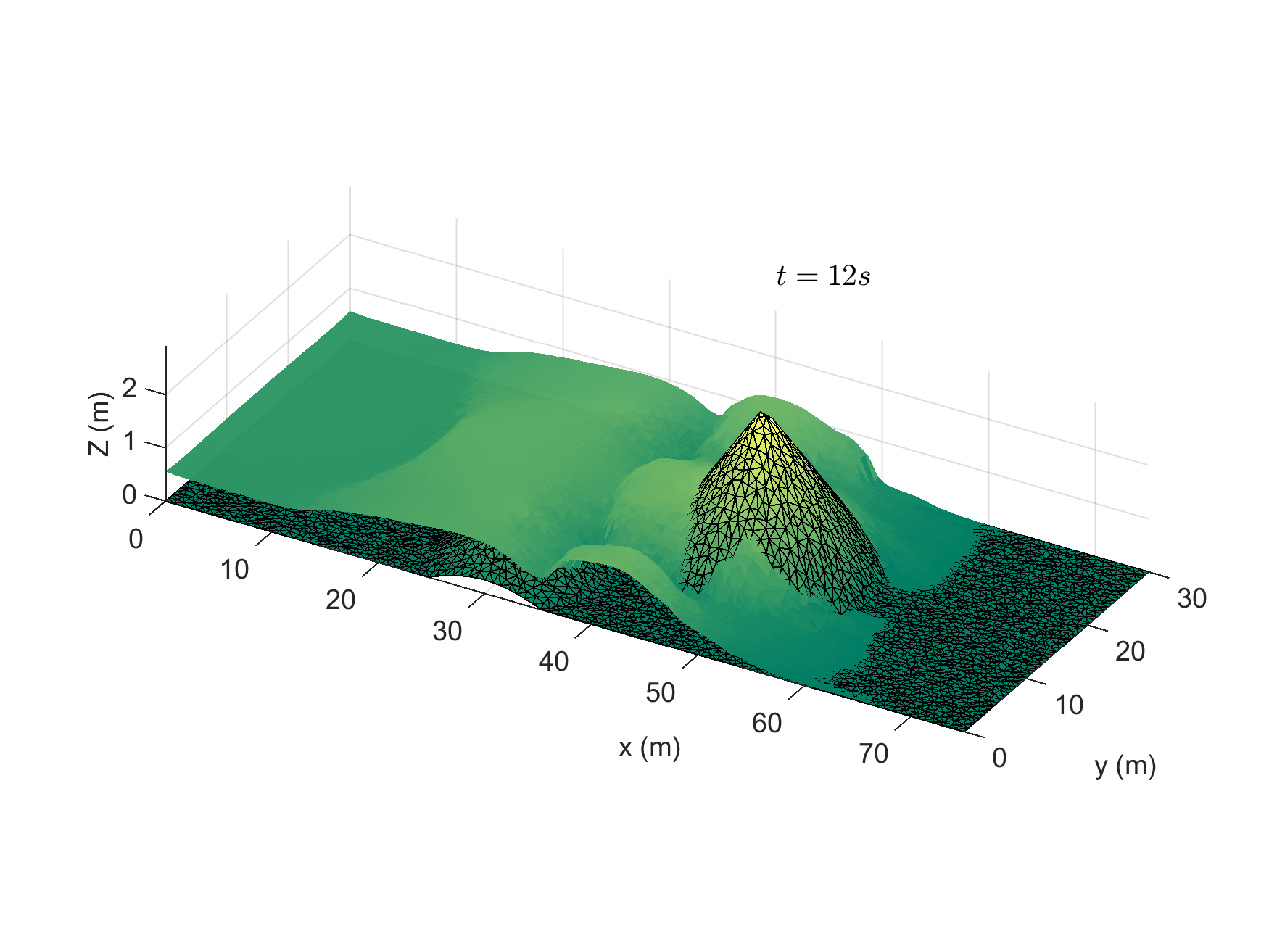}
     \end{minipage}}\\[-12ex]
     \hfill
     \subfloat{
     \begin{minipage}[b]{0.5\textwidth}
         \centering
         \includegraphics[width=\textwidth]{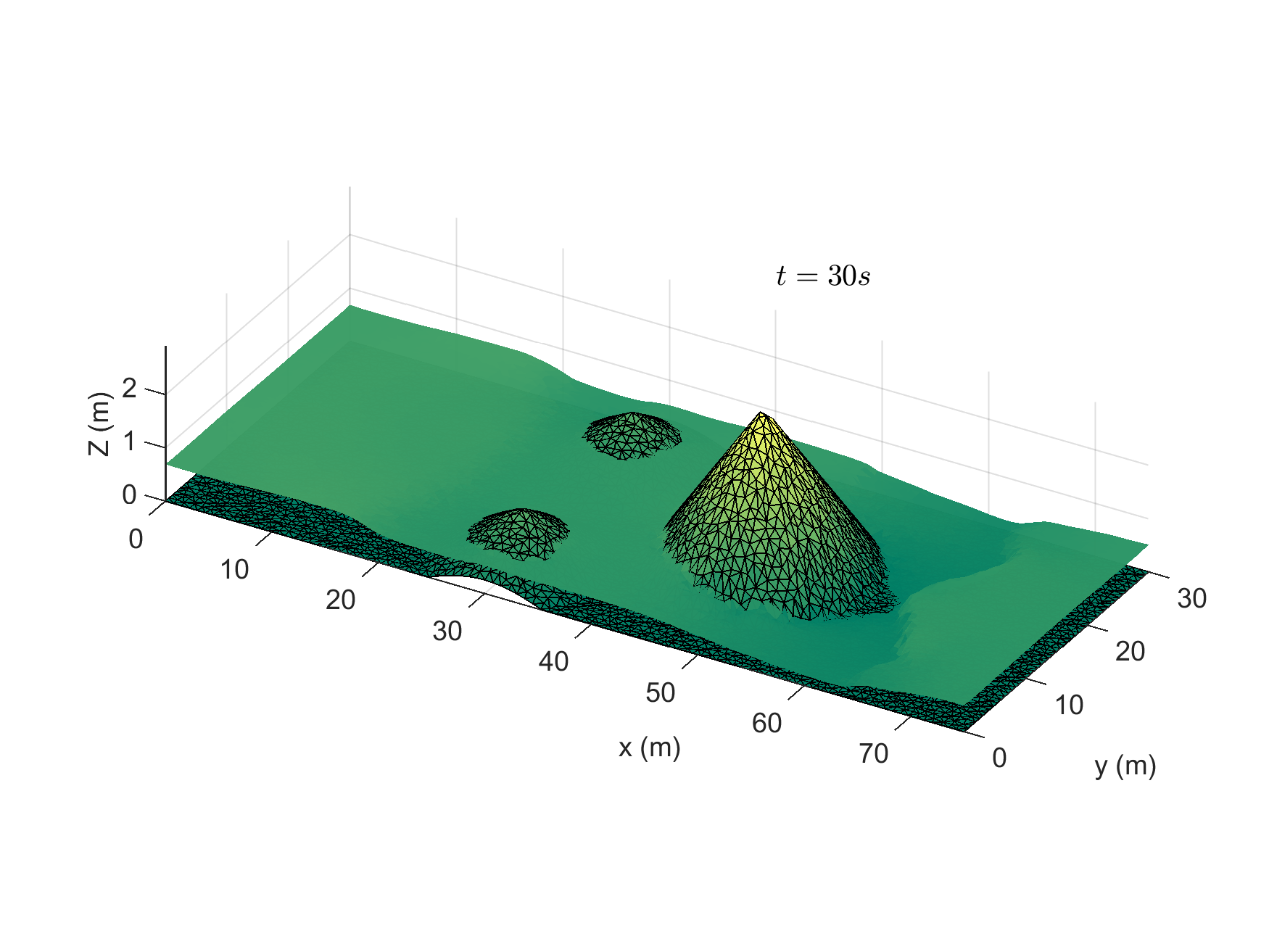}
     \end{minipage}}
     \subfloat{
     \begin{minipage}[b]{0.5\textwidth}
         \centering
         \includegraphics[width=\textwidth]{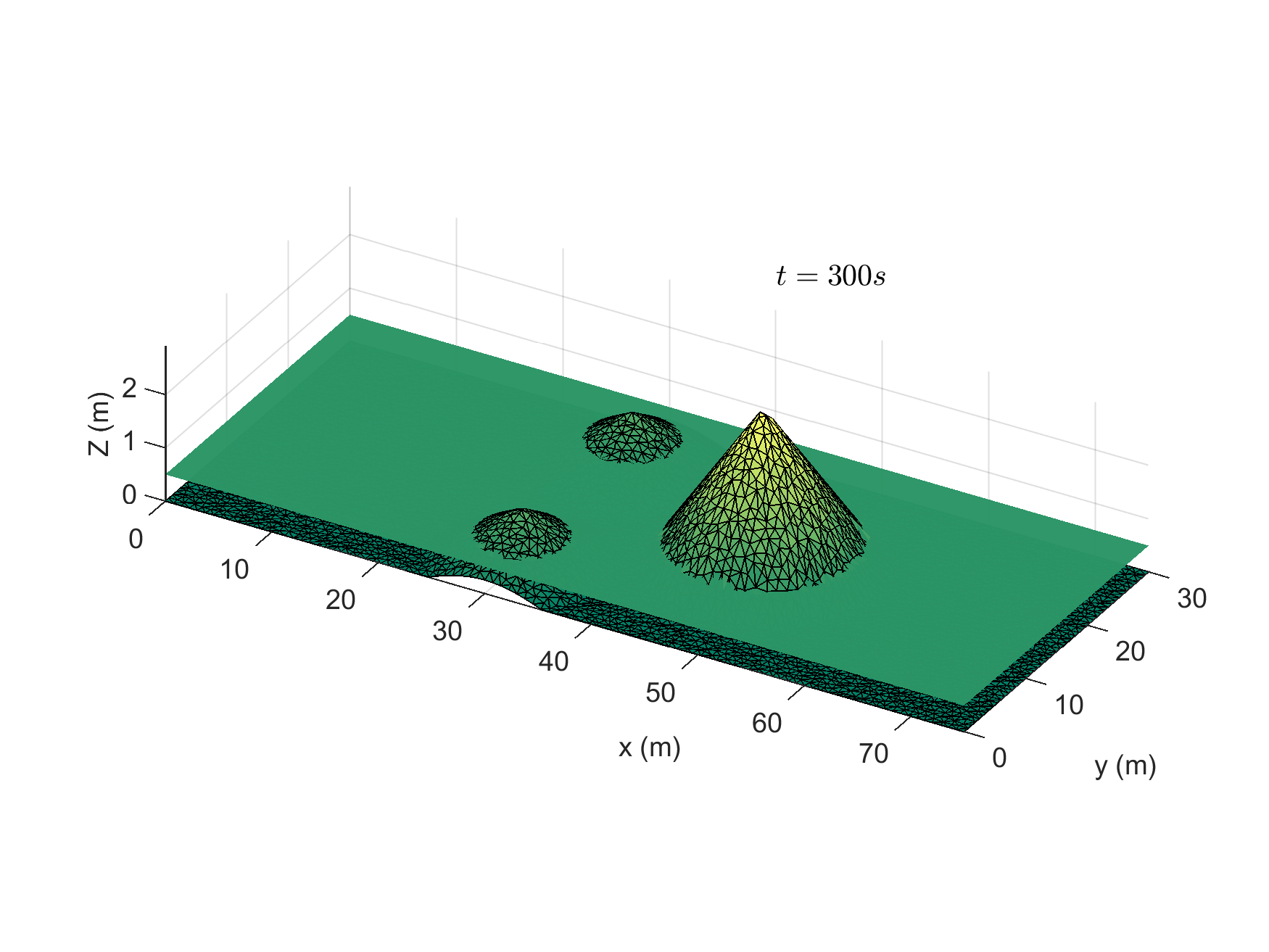}
     \end{minipage}}
        \caption{Dam break wave propagating over three humps: 3D water surface elevation at times t=0,2,6,12,30,300s.}
        \label{fig:ThreeHumps_surface_Levels}
\end{figure}

The computational domain is represented by 5151 meshless points. The dam is located at $x = 16$ ~m with an initial water depth of 1.875 m, and the rest of the channel is considered dry. \cref{fig:ThreeHumps_Topography} shows the bed topography and the upstream still water retained by the dam. The Manning coefficient is set to be, $n=0.018$ m$^{-1/3}$s. At $t = 0$ s, the dam collapses instantaneously. \cref{fig:ThreeHumps_surface_Levels} shows the three-dimensional (3D) view of the free surface elevation at different times, $t=$ 0, 2, 6, 12, 30 and 300 s. At 2 s, the waterfront reaches two lower humps and begins to flow over them. At 6 s, the small humps are entirely submerged, and the wet/dry front reaches the big hump. The front wave runs up the big hump more than halfway up, while water cascades around the side of the hump. The curved reflection bores continue to move upstream and begin to interact with each other and the side walls of the container. At 12 s, the flood water passes either side of the big hump and starts to flood the lee side of the hump. At 30 s, wave interaction between the hump and the downstream wall is noticed. By 300 s, a steady state is achieved, and the peaks of the smaller humps are no longer submerged. The proposed numerical model accurately simulates the complex wetting and drying process and produces results that are very similar to those of \cite{LIANG_Quadtree, NIKOLOS_Delis, Sont_zhou_yang, Brufau,Delis_Kazolea}. 

\subsection{Laboratory dam-break over a triangular hump}
\begin{figure}[hbt!]
     \centering
     \subfloat[]{
     \begin{minipage}[b]{0.85\textwidth}
         \centering
         \includegraphics[width=\textwidth,valign=t]{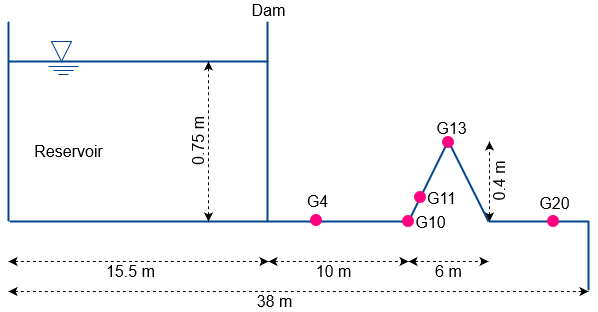}
     \end{minipage}}
        \caption{Laboratory dam-break over a triangular hump: Geometry and gauge locations in the experimental and model setup}
        \label{fig:Laboratory_dam_triangle_hump}
\end{figure}

This test aims to validate the laboratory dam-break flow over a triangular hump as recommended by the EU CADAM project. The physical experiment involves complex hydraulic phenomena, including shocks, transitions between wet and dry beds, and flow over an obstacle. The laboratory setup, illustrated in Fig. \ref{fig:Laboratory_dam_triangle_hump}, includes a reservoir filled with water to a height of $0.75$ m, held back by a dam located at $x = 55.5$ m, with a dry bed downstream within a rectangular channel. A symmetric triangular obstacle, $6$ m long and $0.4$ m high, is positioned in the channel, with its peak located $13$ m downstream the dam. To monitor the flow depth changes, gauges were placed at $4$ m (G4), $10$ m (G10), $11$ m (G11), $13$ m (G13), and $20$ m (G20) from the dam, as shown in the Fig. \ref{fig:Laboratory_dam_triangle_hump}. The gauge at G13 is critical because it is located at the peak of the obstacle. The channel has wall boundaries except for the open outlet, and the Manning's roughness coefficient provided by the experimentalists is $0.0125~\text{m}^\text{-1/3}\text{s}$.

\begin{figure}[hbt!]
     \centering
     \subfloat[]{
     \begin{minipage}[b]{0.50\textwidth}
         \centering
         \includegraphics[width=\textwidth,valign=t]{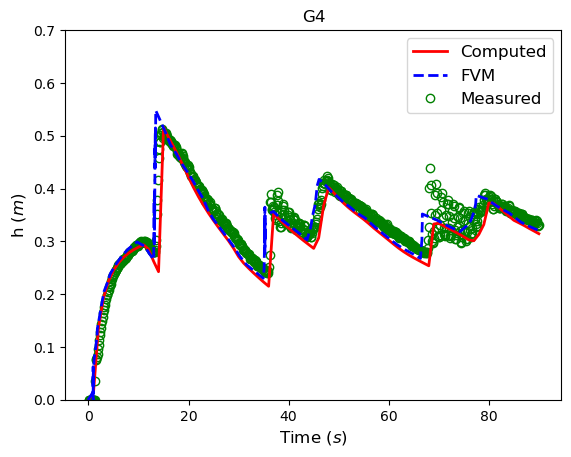}
     \end{minipage}}
     \subfloat[]{
     \begin{minipage}[b]{0.50\textwidth}
         \centering
         \includegraphics[width=\textwidth,valign=t]{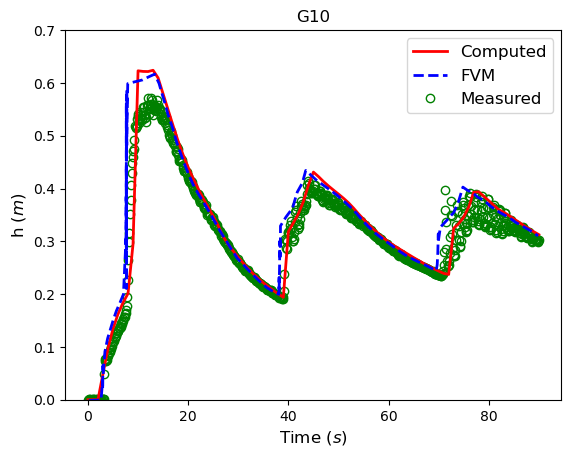}
     \end{minipage}} 
     \hfill
     \subfloat[]{
     \begin{minipage}[b]{0.50\textwidth}
         \centering
         \includegraphics[width=\textwidth,valign=t]{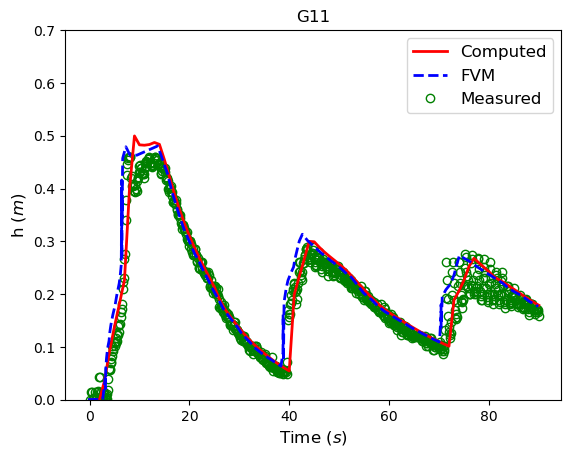}
     \end{minipage}}
     \subfloat[]{
     \begin{minipage}[b]{0.50\textwidth}
         \centering
         \includegraphics[width=\textwidth,valign=t]{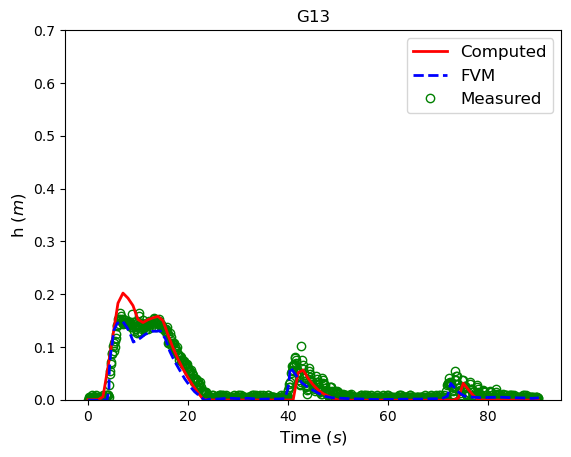}
     \end{minipage}}
     \hfill
     \subfloat[]{
     \begin{minipage}[b]{0.50\textwidth}
         \centering
         \includegraphics[width=\textwidth,valign=t]{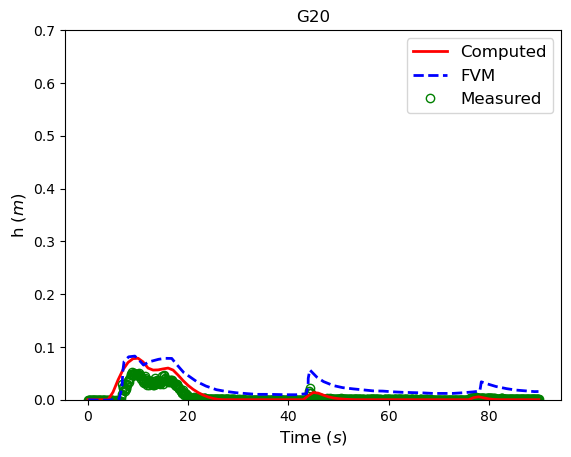}
     \end{minipage}}
        \caption{Laboratory dam-break over a triangular hump: Numerical results are compared with FVM and measured water depth at stations (a) G4, (b) G10, (c) G11, (d) G13, and (e) G20}
        \label{fig:Laboratory_dam_triangle_hump_Gauge_results}
\end{figure}

In the numerical computation, the experimental domain is represented by 7993 irregularly distributed points along the channel. The time step is set to $\Delta t=0.005$ s, with a dry bed tolerance of $h_{tol}=0.000001$ m. The simulation is run for $90$ s. The variation in depths over time at the five gauges are compared with experimental observations, as shown in Fig. \ref{fig:Laboratory_dam_triangle_hump_Gauge_results}. The Root Mean Square Error (RMSE) values for the gauges are computed over a duration of 90 s, with respective values of 0.03534 for G4, 0.03408 for G10, 0.02930 for G11, 0.01620 for G13, and 0.01242 for G20. The simulated flow depths closely match the observed data in terms of depth variation and the arrival time of the waterfront at each gauge. The transitions between wet and dry states at gauge G13 are also accurately predicted. The comparison of flow depth at the peak of the hump shows that the proposed meshless method aligns more accurately with the experimental measurements than the finite volume based solution, reported in previous studies \cite{Kuiry_1D, Liang}. However, as noted in earlier studies \cite{RBF_Triangle_flow,Brufau}, there is a significant deviation from the experimental data at G20, although the flow depth variation at this location throughout the simulation is minimal.

\subsection{Malpasset dam-break}
The Malpasset dam-break event is chosen to apply the proposed meshless model to simulate large-scale real-world application. The dam was located in the Reyran River valley in France. The dam failed in 1959 due to intense rainfall. After the dam failure, maximum water levels along the Reyran River Valley were obtained from a field survey. In 1964, a scaled physical model at a ratio of 1:400 was constructed to investigate the dam-break flow. The peak water level and the arrival time of the flood wave at nine specific points along the valley were recorded in the physical model.

\begin{figure}[hbt!]
     \centering
     \begin{minipage}[b]{0.85\textwidth}
         \centering
         \includegraphics[width=\textwidth,valign=t]{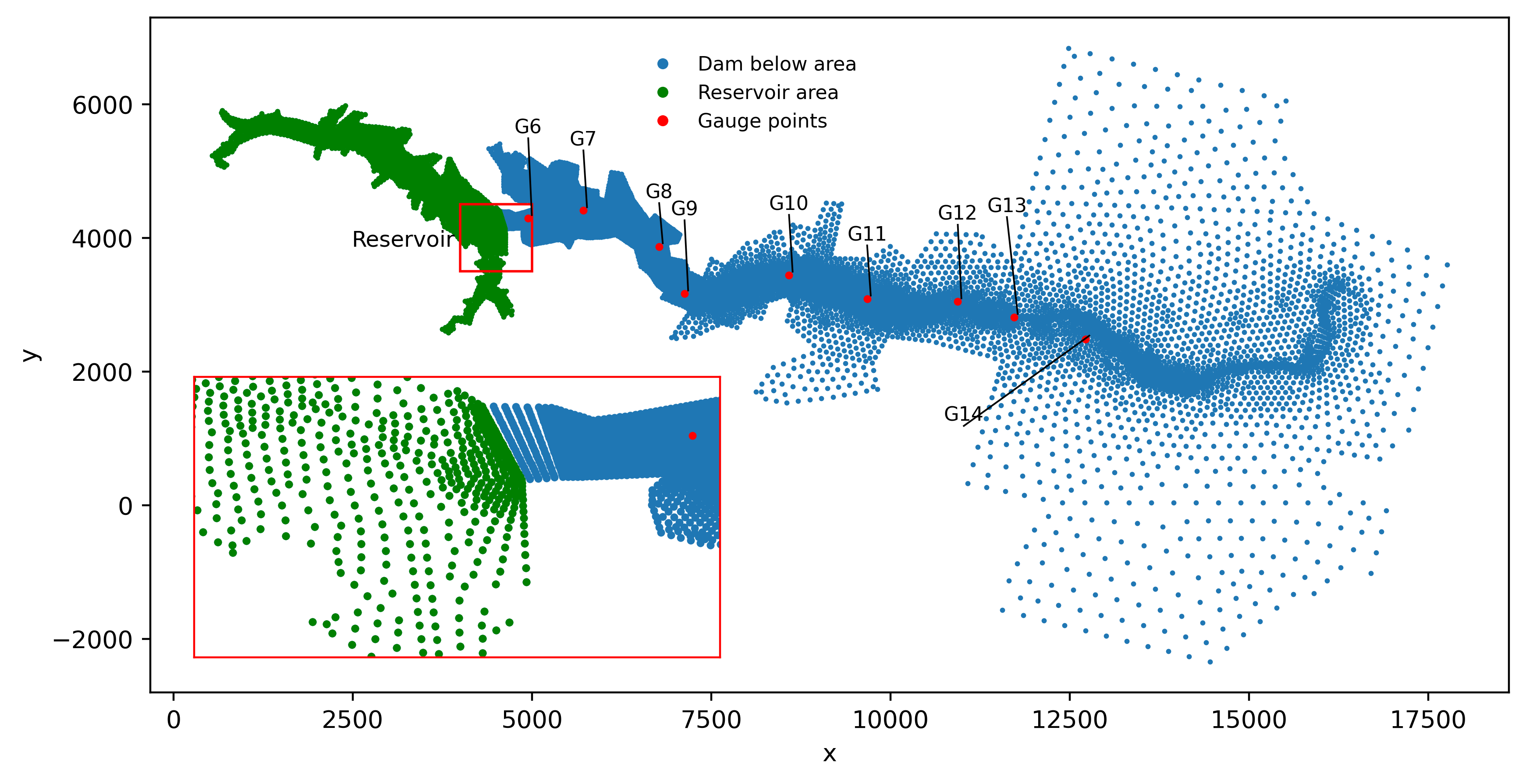}
     \end{minipage}
        \caption{Malpasset dam-break: Point distribution and gauge locations in the dam area}
        \label{fig:Malpasset_Point_dist}
\end{figure}

The computational domain consists of 13,541 points for which topography data was adapted from the CADAM project \cite{Goutal_Maurel_MalpassetDam}. The computational domain and gauge points are shown in \cref{fig:Malpasset_Point_dist}. The dam is considered a straight line between the coordinates (4701.18 m, 4143.41 m) and (4655.50 m, 4392.10 m). The initial water level inside the reservoir is considered to be at 100 m above sea level, and the floodplain is assumed to be initially dry. The boundary conditions are applied as solid walls everywhere except open boundaries near the open sea. Manning's coefficient is 0.025 m$^{-1/3}$s. 

\begin{figure}[hbt!]
     \centering
     \subfloat[]{
     \begin{minipage}[b]{0.50\textwidth}
         \centering
         \includegraphics[width=\textwidth,valign=t]{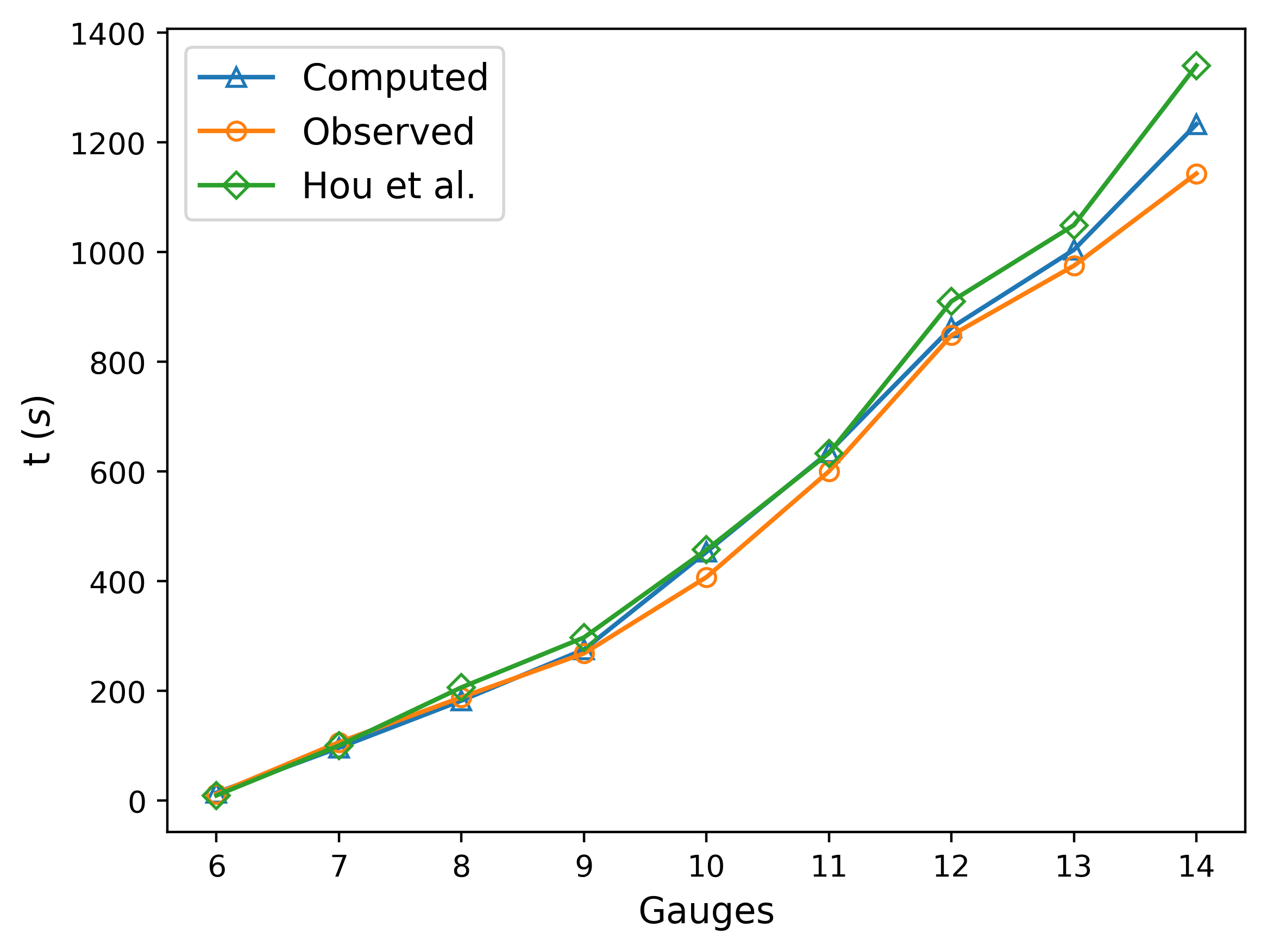}
     \end{minipage}}
     \subfloat[]{
     \begin{minipage}[b]{0.50\textwidth}
         \centering
         \includegraphics[width=\textwidth,valign=t]{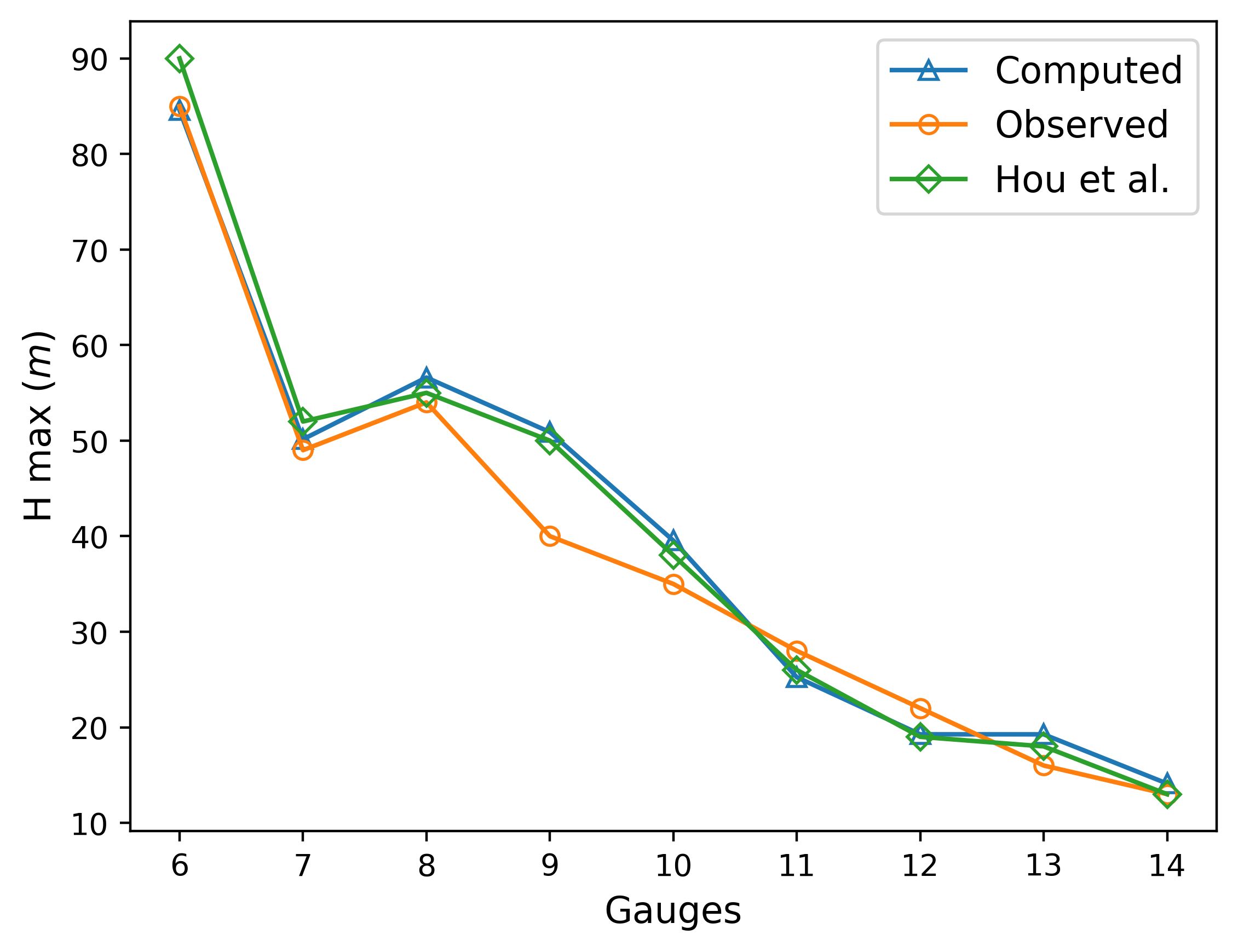}
     \end{minipage}} 
        \caption{Malpasset dam-break: Comparison of numerical results with experimental measurements at gauges (a) arrival times  (b) maximum water levels}
        \label{fig:Malpasset_plots}
\end{figure}

\begin{figure}[hbt!]
     \centering
     \subfloat[]{
     \begin{minipage}[b]{0.8\textwidth}
         \centering
         \includegraphics[width=\textwidth,valign=t]{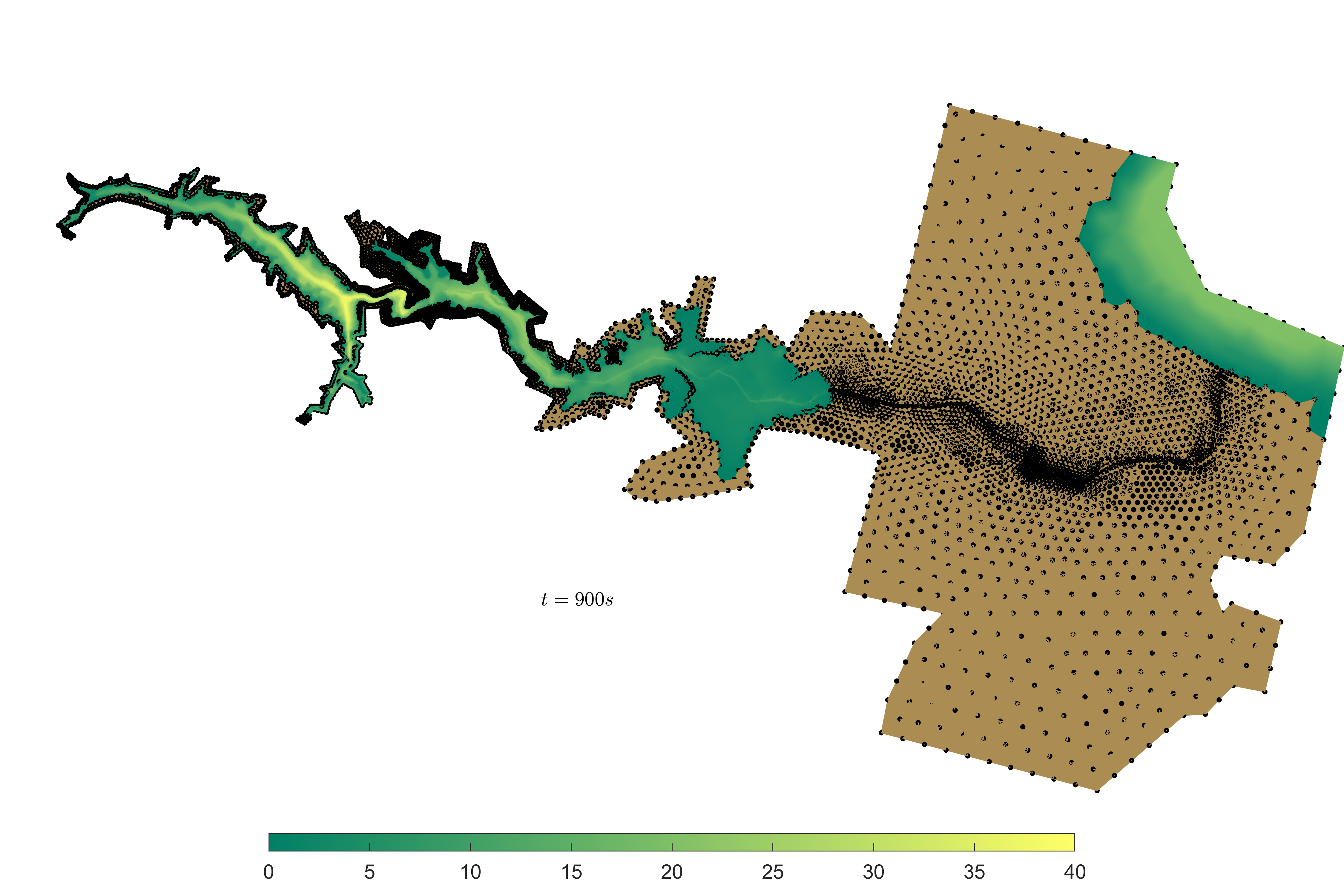}
     \end{minipage}}
     \hfill 
          \subfloat[]{
     \begin{minipage}[b]{0.8\textwidth}
         \centering
         \includegraphics[width=\textwidth,valign=t]{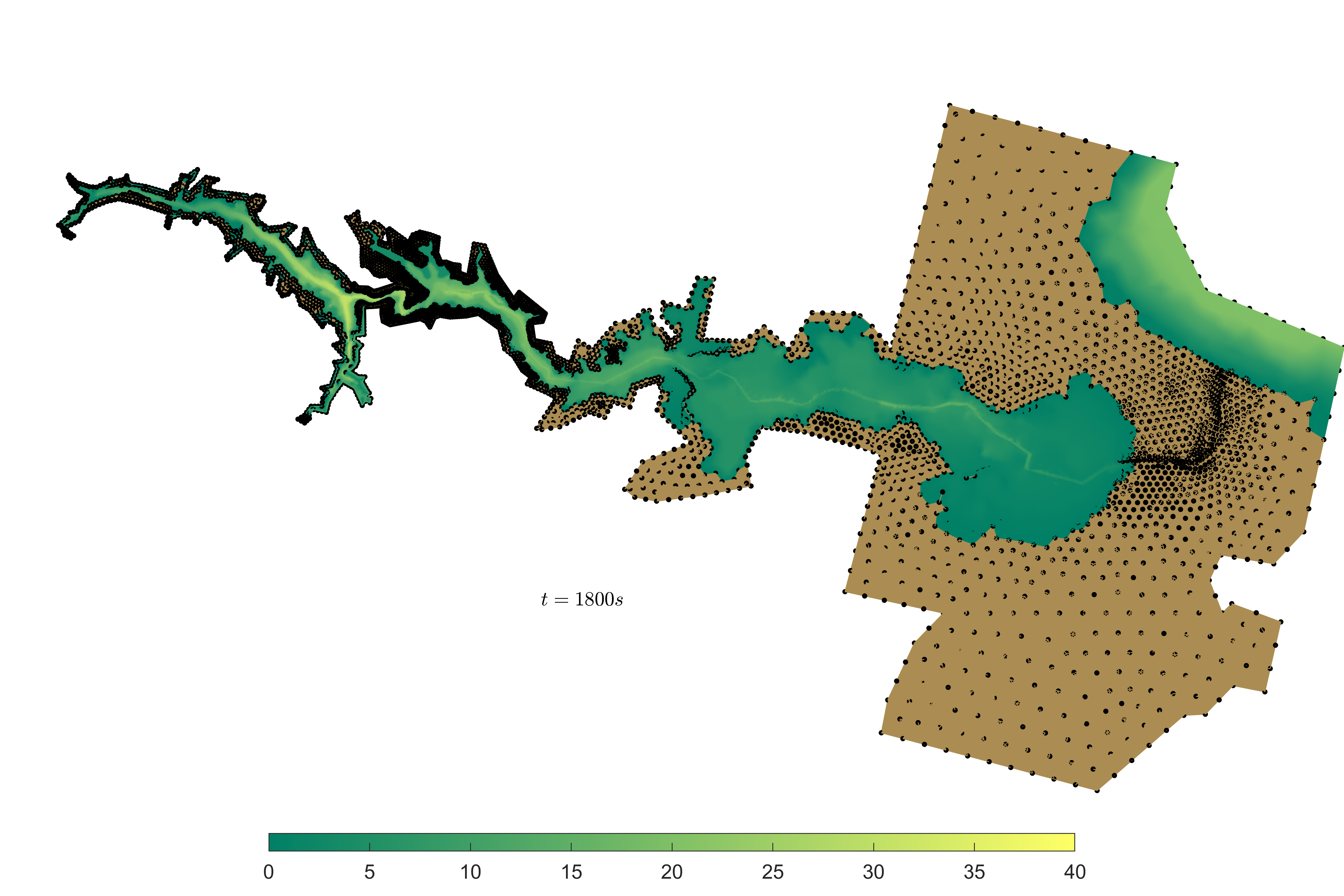}
     \end{minipage}} 
        \caption{Malpasset dam-break: Flood inundation levels (a) at $t = 900s$ (b) at $t = 1800s$}
        \label{fig:Malpasset_flood_levels}
\end{figure}

The simulation is executed until the time $t=3500$ s. \cref{fig:Malpasset_plots} shows the computed flood wave arrival times and the maximum water level comparison at the experimental gauge points of the physical model. The present model shows good agreement between computed results and experimental observations. In addition, similar results can be seen in \cite{Brufau_Navarro, Delis_Argiris}, where the same domain points were used. \cref{fig:Malpasset_flood_levels}(a) and (b) illustrate the progression of the flood wave at two specific time instants, 900 s and 1800 s. At 900 s, the initial stages of the flood wave movement can be observed. By 1800 s, the wave has advanced significantly, reaching the downstream floodplain and demonstrating the extent of its spread over time. A comparable behaviour can be found in \cite{Ying_Wang_2009}.

\section{Conclusions}
A meshless geometric conservation weighted least square (GC-WLS) method with an approximate Riemann solver (HLL) for capturing shocks is developed to solve the 2D shallow water equations. In the proposed method, a collection of irregularly scattered points describes the physical domain. A local cluster of nearby points, termed satellites, is identified for each point. The formulation of local approximations for spatial derivatives relies on meshless coefficients. These coefficients are determined by applying the geometric conservation law and first-order consistency through the Lagrange multiplier method, ensuring discrete local conservation within the approach. Further, the meshless coefficients are utilized to compute convective fluxes using the HLL Riemann solver and source terms. The model automatically detects the wet-dry interface and numerically handles the wetting and drying phenomenon by locally adjusting the water surface elevation. The developed meshless model is verified and validated with several test cases. The simulated water surface elevation and velocity variations for an ideal dam-break on the wet and dry beds are in close agreement with the analytical solutions. The developed method accurately captures flow in a 2D frictional parabolic bowl. The model also captures the dynamic flood process on topography with three humps. The wave propagation due to a dam-break on a dry bed over a triangular obstacle is well simulated, and the computed results closely follow the experimental observations. Also, the shifting of dry and wet conditions is well captured. The proposed meshless method is found to be accurate and can capture shocks and flow discontinuity.










\printcredits

\section*{Declaration of competing interest}
The authors declare that they have no known competing financial
interests or personal relationships that could have appeared to
influence the work reported in this paper.

\bibliographystyle{cas-model2-names}

\bibliography{cas-refs.bib}



\end{document}